\documentclass{article}
\usepackage{amsfonts,amsmath,amssymb,authblk,floatrow,fullpage,graphicx,tabularx,threeparttable,tikz,pgfplots,xcolor,lscape,multirow,verbatim, algorithm, algpseudocode}
\usepackage[colorlinks]{hyperref}
\usepackage[noabbrev,capitalise,sort&compress]{cleveref}
\usepackage{caption,subcaption}
\usepackage{appendix}
\usepackage[textsize = small]{todonotes}
\usepackage[nohyperlinks,nolist]{acronym}
\usepackage{import}
\usepackage{makecell}
\usepackage{placeins}
\usetikzlibrary{arrows,arrows.meta,calc,fit,spy,patterns}

\renewcommand{\vec}[1]{\boldsymbol{#1}}         
\newcommand{\mat}[1]{\boldsymbol{#1}}           
\newcommand{\parder}[2]{\frac{\partial #1}{\partial #2}}

\newcommand{\absder}[2]{\frac{\text{d} #1}{\text{d} #2}}
\definecolor{ccao}{rgb}{0.596, 0.251, 0.043}
\definecolor{ccaa}{rgb}{0.314, 0.882, 0.957}
\definecolor{ael}{rgb}{0.063, 0.714, 0.188}
\definecolor{elel}{rgb}{0.478, 0.953. 0.373}
\definecolor{elc}{rgb}{0.427, 1, 0.46}
\definecolor{cccc}{rgb}{0.796, 0.145, 0.129}
\definecolor{elccc}{rgb}{0.216, 0.114, 0.781}
\definecolor{ccco}{rgb}{0.533, 0.322, 0.055}
\newfloatcommand{capbtabbox}{table}[][\FBwidth]
\pgfplotsset{compat=newest}
\begin{document}
\title{Anaysis of the validity of P2D models for solid-state batteries in a large parameter range}
\author{Stephan Sinzig$^{1,2}$, Christoph P. Schmidt$^{1}$, Wolfgang A. Wall$^{1,2}$}
\affil{\small $^1$ Technical University of Munich, Germany, TUM School of Engineering and Design, Institute for Computational Mechanics, Boltzmannstra\ss e 15, 85748 Garching bei M\"unchen \\ $^2$ TUMint.Energy Research GmbH, Lichtenbergstra\ss e 4, 85748 Garching bei M\"unchen, Germany}

\date{}
\maketitle
\section*{Abstract}
Simulation models are nowadays indispensable to efficiently assess or optimize novel battery cell concepts during the development process. Electro-chemo-mechano models are widely used to investigate solid-state batteries during cycling and allow the prediction of the dependence of design parameters like material properties, geometric properties, or operating conditions on output quantities like the state of charge. One possibility of classification of these physics-based models is their level of geometric resolution, including three-dimensionally resolved models and geometrically homogenized models, known as Doyle-Fuller-Newman or pseudo two-dimensional models. Within this study, the advantages and drawbacks of these two types of models are identified within a wide range of the design parameter values. Therefore, the sensitivity of an output quantity of the models on one or a combination of parameters is compared. In particular, the global sensitivity, i.e., the sensitivity in a wide range of parameter values, is computed by using the Sobol indices as a measure. Furthermore, the local sensitivity of the difference in the output quantities of both models is evaluated to identify regions of parameter values in which they contain significant deviations. Finally, remarks on the potential interplay between both models to obtain fast and reliable results are given.
\section{Introduction}
Simulation models play an important role during the development process of novel batteries. An appropriate computational model enables, for example, the efficient identification of the optimal microstructure within a battery cell, an optimal combination of materials, or limits of extreme cycling scenarios and their combinations. All this may be done before the cell is manufactured for the first time and thereby drastically speed up the development time and reduce the development cost. However, the designer of a model should be aware of \textit{all models are wrong, but some are useful}~\cite{Box1976}, meaning that a model should be defined for a specific research question, e.g., with respect to one particular quantity of interest while being as simple as possible. A model is never suitable to predict the behavior for all research questions that come up during the development of battery cells. Discussions about the generality of the conclusions that can be drawn from simulation models in the context of battery cells are ongoing. For example, some authors suggest that it is difficult to draw general conclusions from complex models~\cite{Bielefeld2023}, others state that the ability to predict the cell behavior of simple models is more limited compared to complex models~\cite{Ramadesigan2012}.\\
Physics-based continuum models that couple phenomena from electrochemistry and solid mechanics are often used to reliably investigate the transport of mass and charge together with the development of stresses and deformations during the cycling of a solid-state battery (SSB) cell.
Within the class of continuum models for SSBs, different types of models exist. One possibility of characterization is the level of geometric resolution. Mainly, three types of continuum models with a different level of geometric resolution exist with no sharp boundaries between them (c.f.~\cite{Zhao2019a,Krewer2018} for an elaborate overview of different models): One type is given by three-dimensionally resolved models in which the complex microstructure is resolved in three dimensions of space~\cite{Schmidt2023,Bucci2017,Bistri2020,Latz2015,Salvadori2015,Bai2021}. Various extensions have been presented to include further physical effects like grain boundaries~\cite{Sinzig2024}, space-charge layers~\cite{Sinzig2023,Neumann2021,Braun2015}, coating layers~\cite{Sinzig2023a,Javed2020}, delamination between the components~\cite{Schmidt2024}, stripping and plating~\cite{Fang2022,Hein2020a}, or SEI formation~\cite{Horstmann2019}. Another type is summarized as single-particle models in which one active material particle is embedded into the solid electrolyte. The third type is summarized as pseudo two-dimensional (P2D) models. Within models of this type, the complex microstructure is geometrically not resolved but considered by effective transport properties that originate from a mathematical homogenization process and reduce the conservation equations to one dimension. This one-dimensional model is pointwise coupled with multiple particles in which the intercalation within the active material is modeled. According to their inventors, these models are synonymously called Doyle-Fuller-Newman models. P2D models are, in general, computationally cheaper compared to three-dimensionally resolved models. However, they lose geometric information during the homogenization process, which might become decisive. Various extensions for P2D models are presented to include further physical effects (c.f.~\cite{Jokar2016} for an overview of P2D models), like the inclusion of local fluctuations within the particles~\cite{Traskunov2022}, the inclusion of a size distribution of the particles~\cite{Kirk2021}, the inclusion of the development of a solid-electrolyte interphase~\cite{Safari2009}, or the inclusion of stripping and plating~\cite{Yang2018b}. In contrast to the inclusion of further physical effects within resolved models, the extensions are often phenomenological to reproduce measured quantities and are not derived from physical principles. Different studies are available to fit the parameters of the P2D model to experimental results~\cite{Khalik2021,Lu2022}. However, studies suggest a critical view on the usage of P2D models when it comes to the quantitative identification of an optimal battery cell~\cite{An2021,Schmidt2021,Goldin2012,Park2015}. \\
We contribute to the ongoing discussion of the required level of geometric complexity by comparing selected model output quantities, like the state of charge for certain cycling scenarios computed with a three-dimensionally resolved continuum model for SSBs, with those from a P2D continuum model. Multiple strategies are followed to assess similarities and differences: The output quantities are compared for a typical battery cell with a typical cycling scenario. Afterwards, the sensitivity of the output quantities on input parameters is computed and compared between the models. In contrast to other studies, the comparison is in a global sense, i.e., in a wide range of input parameters and not at one fixed set of values of the parameters or in its neighborhood, and it considers the interaction between parameters. Based on this, the local sensitivity, i.e., the sensitivity of the output quantity as a function of a parameter value, is compared to identify regions of the parameter values in which the output quantities deviate most. \\
In the following, both models are briefly summarized. Afterwards, the output quantities of both models are analyzed to point out the weaknesses of the individual models. Finally, suggestions for the interplay between both models are given to profit best from the advantages of both models.
\FloatBarrier
\section{Summary of electro-chemo-mechano continuum models for SSBs}
The geometrically fully resolved model and the P2D model are both based on the same conservation equations for mass, charge, and linear momentum to model SSBs.
\subsection{Governing equations - conservation of mass, charge, and linear momentum}
\label{sec:conservation_equations}
\begin{figure}[ht]
  \centering
  \def\svgwidth{0.8\textwidth}
  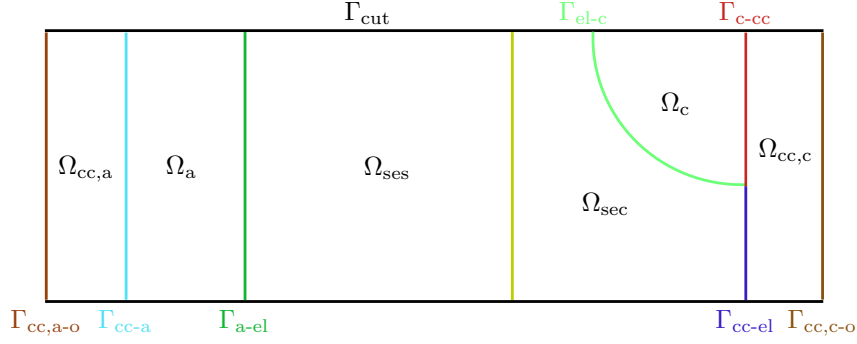
  \caption{Schematic sketch of the components~$\Omega_i$ of an SSB cell, boundaries~$\Gamma_i$ and interfaces between the components~$\Gamma_{i-j}$ }
  \label{fig:schematic_geometry}
\end{figure}

We present a brief summary of the set of conservation equations of mass, charge, and linear momentum for SSBs in a continuous form (for details, see, e.g.,~\cite{Schmidt2023}). The equations are defined within the components of the battery cell in \cref{fig:schematic_geometry}, i.e., the electrodes consisting of the metal anode~$\Omega_\text{a}$ and the cathode active material~$\Omega_\text{c}$ ($\Omega_\text{ed} = \Omega_\text{a} \cup \Omega_\text{c}$), the solid electrolyte separator ($\Omega_\text{ses}$), the solid electrolyte within the composite cathode~$\Omega_\text{sec}$, and the current collectors ($\Omega_\text{cc} = \Omega_\text{cc,a} \cup \Omega_\text{cc,c}$).
The domain of the solid electrolyte is composed of the solid electrolyte separator and the solid electrolyte within the composite cathode ($\Omega_\text{el} = \Omega_\text{ses} \cup \Omega_\text{sec}$). The domain of the composite cathode is defined by~$\Omega_\text{comp c} = \Omega_\text{c} \cup \Omega_\text{sec}$. An interface between the domains~$\Omega_i$ and~$\Omega_j$ is defined as~$\Gamma_{i-j}$. The geometric definition is completed by physically meaningful boundaries~$\Gamma_\text{cc,a-o}$ and~$\Gamma_\text{cc,a-o}$ and modeling boundaries ~$\Gamma_\text{cut}$ required to limit the size of the geometry. \\
The governing equations are
\begin{subequations}
\begingroup
\allowdisplaybreaks
\begin{align}
    \nabla_0 \cdot (\mat{F} \mat{S}) &= \vec{b}_0 \quad \text{in } \Omega_0, \\
    \mat{F} = \nabla \vec{u} + \mat{1} &= \mat{F}_\text{el} \mat{F}_\text{growth}, \\
    \mat{C}_\text{el} &= \mat{F}_\text{el}^\text{T} \mat{F}_\text{el}, \\
    \mat{S} &= 2 \text{det} (\mat{F}_\text{growth}) \mat{F}_\text{growth}^{-1} \parder{\Psi_\text{el}}{\mat{C}_\text{el}} \mat{F}_\text{growth}^\text{-T}, \\
    \mat{F}_\text{growth} &= \mat{1} + \left[p \ \text{det}(\mat{F}) \frac{ \left( n_\text{ed} - n_\text{ed}^0 \right) }{V} \right] \mat{G} \quad \text{in } \Omega_\text{a}, \label{eq:growth_deposition} \\
    \mat{F}_\text{growth} &= \left(\frac{f(\chi) + 1}{f(\chi_0) + 1}\right)^\frac{1}{3} \mat{1}\quad \text{in } \Omega_\text{c}, \\
    \vec{u} \cdot \vec{n} &= 0 \quad \text{on } \Gamma_\text{cc,a-o} \cup \Gamma_\text{cut} \\
    \mat{F} \mat{S} \vec{N} &= \vec{u} k \quad \text{on } \Gamma_\text{cc,c-o} \\
    \absder{c}{t} - \nabla \cdot (D \nabla c) &= 0 \quad \text{in } \Omega_\text{ed}, \\
    \nabla \cdot (- \sigma \nabla \Phi) &= 0 \quad \text{in }  \Omega_\text{ed} \cup \Omega_\text{cc}, \label{eq:laplace_electrode} \\
    \nabla \cdot (- \kappa \nabla \Phi) &= 0 \quad \text{in }  \Omega_\text{el}, \label{eq:laplace_solid_electrolyte}\\
    \Phi &= 0 \quad \text{on} \ \Gamma_\text{cc,a-o}, \\
    - \vec{i} \cdot \vec{n} &= \hat{i} \quad \text{on } \Gamma_\text{cc,c-o}, \\
    \vec{i} \cdot \vec{n} = \vec{j} \cdot \vec{n} &= 0 \quad \text{on } \Gamma_\text{cut}, \\
    i = \vec{i}_\text{ed} \cdot \vec{n}_\text{ed} = - \vec{i}_\text{el} \cdot \vec{n}_\text{el} &=  i_0 \left[\text{exp} \left(\frac{\alpha_\text{a} F \eta}{R T}\right) - \text{exp} \left(\frac{-(1-\alpha_\text{a}) F \eta}{R T}\right) \right] \quad \text{on } \Gamma_\text{el-ed} \label{eq:butler_volmer} \\
    \vec{j}_\text{ed} \cdot \vec{n}_\text{ed} &= - \vec{j}_\text{el} \cdot \vec{n}_\text{el} = \frac{F z}{t_+} i \quad \text{on } \Gamma_\text{el-ed} \\
    \vec{i}_\text{ed} \cdot \vec{n}_\text{ed} &= - \vec{i}_\text{cc} \cdot \vec{n}_\text{cc}  = \frac{\Phi_\text{cc} - \Phi_\text{ed}}{r_\text{i}} \quad \text{on } \Gamma_\text{ed-cc},\\
    c(t=0) &= 
    \begin{cases}
        c_{0,\text{a}} \quad & \text{in } \Omega_\text{a} \\
        c_{0,\text{c}} \quad & \text{in } \Omega_\text{c}
    \end{cases},
\end{align}
\endgroup
\end{subequations}
with the overvoltage $\eta = \Phi_\text{ed} - \Phi_\text{el} - \Phi_0(c)$ at the interface~$\Gamma_\text{ed-el}$. The used symbols are listed in \cref{table:list_of_symbols}.
\begin{table}[ht]
    \begin{tabular}{c | l}
        \textbf{symbol} & \textbf{description} \\
        \hline
        $\vec{b}_0$ & volumetric load in initial configuration\\
        $c$ & concentration of species in the electrode \\
        $D$ & diffusion coefficient in the electrode \\
        $F$ & Faraday constant \\
        $f$ / $p$ & relative volume change / growth factor \\
        $\mat{F}$ & deformation gradient. Subscript 'el' denotes the elastic part and 'growth' the growing part \\
        $\mat{G}$ & tensor with direction of growth \\
        $\vec{i}$ & electric current density \\
        $i_0$ & exchange current density \\
        $\vec{j}$ & mass flux density \\
        $k$ & external stiffness \\
        $n$ & amount of substance \\
        $\vec{n}, \vec{N}$ & normal vector on boundary $\Gamma$ \\
        $R$ & universal gas constant \\
        $r_\text{i}$ & interface resistance \\
        $\mat{S}$ & second Piola-Kirchhoff stress \\
        $T$ & temperature \\
        $t$ & time \\
        $t_+$ & transference number in the solid electrolyte \\
        $\vec{u}$ & displacement \\
        $V$ & volume \\
        $z$ & charge number \\
        $\alpha_\text{a}$ & anodic symmetry coefficient \\
        $\kappa$ & ionic conductivity in the solid electrolyte \\ 
        $\Phi$ & electric potential \\
        $\Phi_0$ & equilibrium potential \\
        $\Psi_\text{el}$ & strain energy function \\
        $\sigma$ & electronic conductivity in the electrodes and current collectors \\
        $\Gamma$ & boundary \\
        $\Omega$ & domain \\   
        \hline
    \end{tabular}
    \caption{List of used symbols.}
    \label{table:list_of_symbols}
\end{table}
The operator~$\nabla_0$ denotes the Nabla operator evaluated in the initial configuration.
\subsection{Geometrically resolved model}
Within a geometrically resolved model, the equations in \cref{sec:conservation_equations} are solved in three dimensions without modification as all components of the SSB cell are geometrically resolved. In this study, an artificially generated representation of the microstructure is used as also done, for example, in~\cite{Schmidt2023,Sinzig2023a}. The equations are discretized in space using the finite element method and in time using the one-step theta method. The resulting system of equations is solved in a monolithic fashion.
\subsection{Geometrically homogenized P2D model}
P2D models are based on the work of Newman and coworkers~\cite{Doyle1992,Doyle1995}. By a homogenization operator, the conservation equations in \cref{sec:conservation_equations} are reduced to one-dimensional equations, which account for the complex microstructure by effective material parameters. \\
In this study, the homogenization follows the methodologies outlined in~\cite{Goldin2012,Pereira2022}. The entire domain is split into~$i$ non-overlapping subdomains $\Omega = \sum_i \Omega_i$ representing the different components of the SSB cell. A quantity~$\Phi$ is non-zero in one of the subdomains~$\Omega_\text{phase}$ and zero in the other domains. The following homogenization operations are defined
\begin{align}
  \bar{\phi} &= \frac{1}{V} \int_\Omega \Phi \text{ d} \Omega = \frac{1}{V} \int_{\Omega_\text{phase}} \Phi \text{ d} \Omega + \underbrace{\frac{1}{V} \int_{\Omega \backslash \Omega_\text{phase}} \Phi \text{ d} \Omega}_{=0}, \label{eq:assb_model:newman_average1}\\
  \bar{\phi}_\text{phase} &= \frac{1}{V_\text{phase}} \int_{\Omega_\text{phase}} \Phi \text{ d} \Omega. \label{eq:assb_model:newman_average2}
\end{align}
All homogenized quantities are indicated with a bar in the following. The share of the volume of one subdomain $\Omega_i$ of the entire domain is defined as $\epsilon_i = \frac{V_i}{V}$ which gives a relation between the \cref{eq:assb_model:newman_average1,eq:assb_model:newman_average2} by $\bar{\phi} = \epsilon_\text{phase} \bar{\phi}_\text{phase}$. \\
For the homogenization of the divergence of a flux~$\vec{N}$, the following relation is used
\begin{equation}
  \overline{\nabla \cdot \vec{N}} = \epsilon \nabla \cdot \bar{N} + \frac{1}{V} \int_\Gamma N_\text{i} \text{ d} \Gamma = \epsilon \nabla \cdot \bar{N} + A \bar{N}_\text{i},
\end{equation}
to ensure the conservation of species within a reference domain~$\Omega$. This homogenization accounts for a potential interaction in terms of an exchange flux~$\bar{N}_\text{i}$ with another phase across the boundary~$\Gamma$ within the domain~$\Omega$. The prefactor~$A$ is the specific surface area, i.e., the interface area to the other phase within the domain~$\Omega$. It is given by $A=\epsilon_i \frac{A_\text{b}}{V_\text{b}}$ with the surface area~$A_\text{b}$ and the volume~$V_\text{b}$ of the domain of the other phase. For spheres with diameter~$d_\text{b}$ as the other phase, the prefactor is~$A = 6 \frac{\epsilon_i}{d_\text{b}}$. The homogenized flux~$\bar{N}$ is defined in terms of the tortuosity~$\tau$, i.e., $\bar{N} = \frac{N}{\tau}$, which is often approximated by the Bruggeman equation
\begin{equation}
  \tau = \epsilon^b,
  \label{eq:assb_model_bruggeman}
\end{equation}
where the exponent~$b$ is often chosen to $b=-0.5$~\cite{Goldin2012}. \\
The conservation of mass and charge is solved within the composite cathode~$\Omega_\text{comp c}$, the anode~$\Omega_\text{a}$, and the solid electrolyte separator~$\Omega_\text{ses}$. The composite cathode consists of two phases, i.e., the cathode active material and the solid electrolyte, while the other two domains are modeled as single phases, assuming the absence of voids and further additives. Within the single-phase domains, the conservation equations are formulated one-dimensionally
\begin{align}
  \parder{(\sigma \parder{\Phi}{x})}{x} &= 0 \quad \text{in } \Omega_\text{a} \\
  \absder{c}{t} - \parder{(D \parder{c}{x})}{x} &= 0 \quad \text{in } \Omega_\text{a} \\
  \nabla \cdot (\kappa \nabla \Phi) &= 0 \quad \text{in } \Omega_\text{ses}, \\
  \parder{\left(F S\right)}{x} + b_0 &= 0 \quad \text{in } {\Omega_0}_\text{a} \cup {\Omega_0}_\text{ses}.
\end{align}
Within the composite cathode, the conservation of charge is computed for the two phases by the outlined homogenization strategy
\begin{align}
  \parder{\left( \frac{\epsilon_\text{el}}{\tau_\text{el}}\kappa \parder{\bar{\Phi}_\text{el}(x)}{x} \right)}{x} + A_\text{el-c} \bar{N}_{q_\text{el-c}} &= 0  \quad \text{in} \ \Omega_\text{comp c}, \label{eq:homogenized_electrolyte} \\
  \parder{\left( \frac{\epsilon_\text{c}}{\tau_\text{c}}\sigma \parder{\bar{\Phi}_\text{c}(x)}{x} \right)}{x} + A_\text{el-c} \bar{N}_{q_\text{el-c}} &= 0  \quad \text{in} \ \Omega_\text{comp c}, \label{eq:homogenized_electrode} \\
  \parder{\left(\bar{F} \bar{S}\right)}{x} + b_0 &= 0 \quad \text{in } {\Omega_0}_\text{comp c}.
\end{align}
The conservation of mass is modeled within equally distributed spheres inside the composite cathode. Due to the homogenization, a sphere is assumed at every point~$x$ in which the conservation of mass is solved
\begin{equation}
    \parder{c(x,r)}{t} - \frac{1}{r^2}\parder{\left(r^2 D \parder{c(x,r)}{r}\right)}{r} = 0.
    \label{eq:homogenized_conservation_mass}
\end{equation}
The conservation of mass within the spheres is coupled with the homogenized equations (\cref{eq:homogenized_electrolyte,eq:homogenized_electrode}) by
\begin{equation}
  D \parder{c}{r} = - \bar{N}_{c_\text{el-c}} = - \frac{\bar{N}_{q_\text{el-c}}}{F z} \quad \text{at} \ r = r_\text{ed}.
  \label{eq:homogenized_coupling}
\end{equation}
The exchanged current~$\bar{N}_{q_\text{el-c}}$ between the solid electrolyte and the electrodes follows the Butler-Volmer equation (c.f. \cref{eq:butler_volmer}). The homogenized elastic constants and the homogenized growth law for the composite cathode are obtained by assuming two extreme cases of composite cathodes: A parallel alignment and a serial alignment of the components (c.f. \cref{sec:averaging_mechanics} for the derivation). \\
The set of equations is again discretized in space using the finite-element method and in time using the one-step theta method. The resulting system of equations is solved in a sequential manner, i.e., the conservation of mass and charge is solved before the conservation of linear momentum.
\FloatBarrier
\section{Comparison of the cycling behavior computed with models based on different geometric resolution}
The cycling behavior computed with a model in which the geometric microstructure is fully resolved is compared to the cycling behavior computed with a P2D model in which the geometry is homogenized. Two strategies are followed for this comparison. First, the evolution of quantities over time and at characteristic points within the microstructure are compared for one specific setup. Second, the models are investigated for their ability to predict the cycling behavior within a wide range of values of the input parameters. The second comparison is motivated by the required ability to reliably predict the cycling behavior for parameter combinations for which neither experimental nor other data is available, which is crucial when, for example, designing or optimizing a novel battery cell. The assumption for these studies is that the resolved model captures the reality more accurately than the P2D model, as both models are derived from the same physical principles, but the P2D model contains further simplification assumptions on the geometry. Note that this does not mean that the resolved model is the more suitable model, as we always aim at the simplest model that can answer the specific research question at hand. Before the cycling behavior is compared, a potential target geometry of a microstructure and the derived homogenized values are summarized together with the definition of the materials and cycling scenarios.
\subsection{Geometry, materials, and cycling scenarios}
\begin{figure}[ht]
  \begin{floatrow}
  \ffigbox{
    \def\svgwidth{0.25\textwidth}
    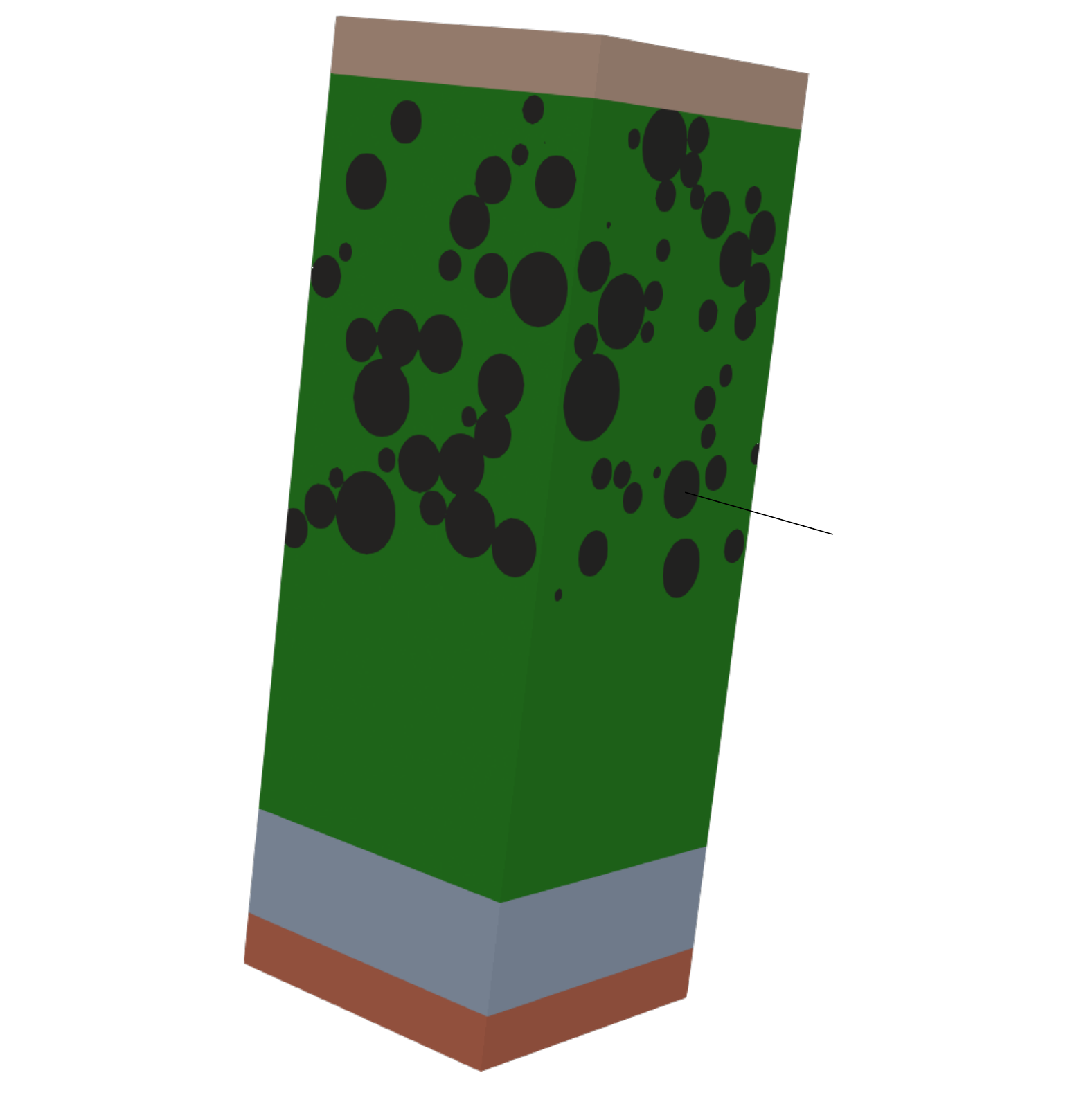
  }{
    \caption{Three-dimensional representation of the computational domain of the SSB cell. The cathode active material is approximated by spherical particles, the anode as a planar metal anode, and the current collectors as a planar foil.  The solid electrolyte fills the voids.} 
    \label{fig:3D_geometry_comparison}
  }
  \capbtabbox{
    \begin{tabular}{| c | c | c |}
    \hline
    \textbf{quantity}                                          & \textbf{value} \\
    \hline
    length of current collectors ($l_\text{cc}$)               & $10 \ \mu \text{m}$ \\
    \makecell[c]{length of \\ composite cathode ($l_\text{comp c}$)} &$85 \ \mu \text{m}$ \\
    length of separator ($l_\text{ses}$)                       & $50 \ \mu \text{m}$ \\
    length of anode ($l_\text{a}$)                             & $20.4 \ \mu \text{m}$ \\
    lateral length ($l_\text{l}$)                              & $60 \ \mu \text{m}$ \\
    \makecell[c]{log-normal distribution of \\ diameter of cathode particles}  & \begin{tabular}{c} $\mu = 1.8189$ \\ $\sigma = 0.4589$ \end{tabular} \\
    \makecell[c]{volumetric ratio of AM \\and SE in composite cathode}       & $\frac{V_\text{AM}}{V_\text{AM} + V_\text{SE}} = 0.4$ \\
    porosity                                                 & $0.2$ \\
    \hline
\end{tabular}
  }{
    \caption{Geometric parameters of a target SSB cell.}
    \label{table:geometric_parameters_comparison}
  }
  \end{floatrow}
\end{figure}

The geometry of the microstructure of the SSB cell is selected to represent the potential microstructure of a target SSB cell. The defining parameters are summarized in \cref{table:geometric_parameters_comparison}.
The artificially generated microstructure required for the solution of the resolved model is shown in \cref{fig:3D_geometry_comparison}. Note that the lateral length, i.e., the size in the x- and y-direction, is artificially reduced compared to a real SSB cell to limit the size of the computational domain.\\
An analysis of the generated microstructure reveals a utilization of $u=0.93$, i.e., the share of the particles that are electronically conductively connected to the current collector. This utilization is not considered within the P2D model as this information is typically not available if the microstructure is not explicitly resolved. The utilization could be included within the P2D model by artificially reducing the available capacity of the cathode by a modification of the specific interface area~$A_\text{el-c}$ if any a priori information on the utilization is available. However, artificially modifying one homogenization parameter might be inconsistent with the other homogenization parameters. \\
The homogenization parameters within the composite cathode required for the P2D model are evaluated as the share of the solid electrolyte~$\epsilon_\text{el} = 0.424$, the share of the active material~$\epsilon_\text{c} = 0.376$, the tortuosity based on the Bruggeman equation within the solid electrolyte~$\tau_\text{el} = 1.54$, and within the active material~$\tau_\text{el} = 1.63$ the mean diameter of the active material particles of $\bar{d} = 6.85 \, \mu \text{m}$, and the specific area of $A_\text{el-c}=0.329 \frac{\mu \text{m}^2}{\mu \text{m}^3}$. Here, the specific area is slightly changed to $A_\text{el-c}=0.344 \frac{\mu \text{m}^2}{\mu \text{m}^3}$ such that the capacity of the SSB cell in both models is exactly the same. \\
The battery cell consists of a composite cathode with NMC-622 as the active material and LPS as the solid electrolyte. The separator is made of LPS, and the anode is made of lithium. All material parameters of the individual components are summarized in \cref{table:material_parameters} in the Appendix \ref{sec:material_parameters}. The homogenized material parameters in the composite cathode are computed based on the homogenization parameters and summarized in \cref{table:homogenized_parameters_comparison}.
\begin{table}[ht]
  \renewcommand{\arraystretch}{1.2}
  \centering
  \caption{Homogenized material parameters used within the composite cathode of the P2D model.}
  \begin{tabular}{| c | c | c |}
    \hline
    \textbf{quantity}                          & \textbf{symbol}   & \textbf{value} \\
    \hline
    electronic conductivity in active material & $\bar{\sigma}(\chi)$    & $\frac{\epsilon_\text{c}}{\tau_\text{c}} \sigma(\chi) = 0.23 \, \sigma(\chi)$ \\
    ionic conductivity in solid electrolyte    & $\bar{\kappa}$          & $\frac{\epsilon_\text{el}}{\tau_\text{el}} \kappa = 3.3 \cdot 10^{-3} \frac{\text{S}}{\text{m}}$  \\
    growth factor                              & $\bar{g}$               & $1.25 \cdot 10^{-7} \frac{\text{m}^3}{\text{mol}}$ \\
    Young's modulus                            & $\bar{E}$               & $95.9 \text{ GPa}$ \\
    \hline
\end{tabular}
  \label{table:homogenized_parameters_comparison}
\end{table}
The SSB cell is cycled with a constant c-rate of 0.25 C at an ambient temperature of $T=298 \text{ K}$ within the bounds of the cell voltage of $\Delta \Phi_\text{low} = 3.6 \ \text{V}$ and $\Delta \Phi_\text{high} = 4.2 \ \text{V}$ beginning with the discharge from the fully charged state. Mechanically, the SSB cell is constrained by a frame with the stiffness $k= 500 \frac{\text{MPa}}{\text{mm}}$ to maintain the mechanical pressure during cycling.
\subsection{Comparison of global and local quantities for a fixed parameter set}
\label{sec:comparison_fixed_parameters}
\begin{figure}[ht]
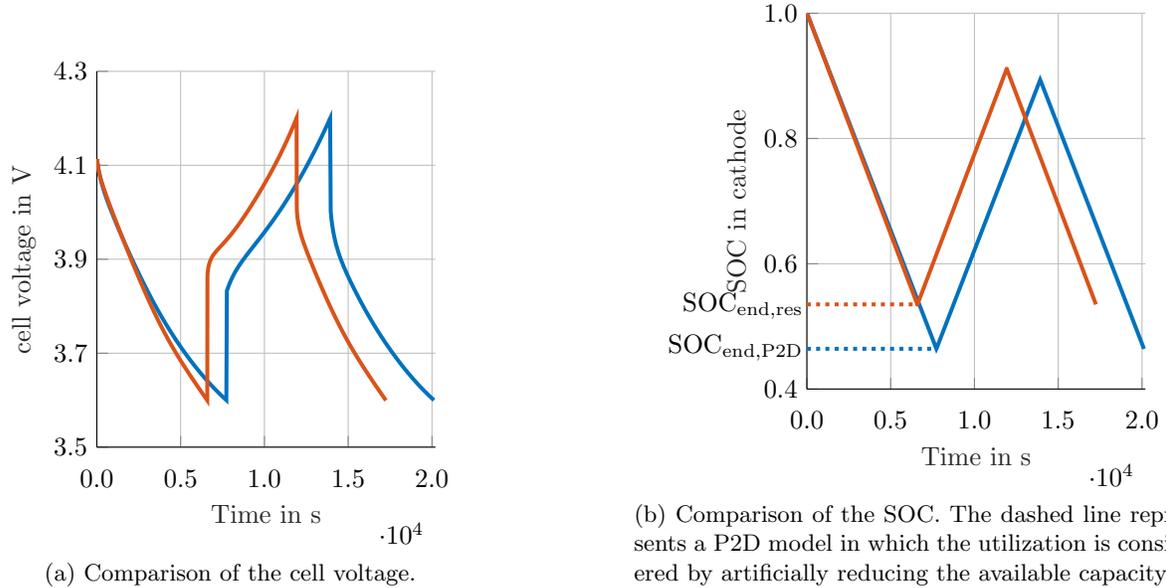

  \centering
  \begin{subfigure}[b]{0.45\textwidth}
    \centering
    \input{figures/comparison_resolved_homogenized_cell_voltage_c_rate.tikz}
    \caption{Comparison of the cell voltage.}
    \label{fig:comparison_resolved_homogenized_cell_voltage_c_rate}
  \end{subfigure}
  \hfill
  \begin{subfigure}[b]{0.45\textwidth}
    \centering
    \input{figures/comparison_resolved_homogenized_soc.tikz}
    \caption{Comparison of the SOC. The dashed line represents a P2D model in which the utilization is considered by artificially reducing the available capacity.}
    \label{fig:comparison_resolved_homogenized_soc}
  \end{subfigure}
  \caption{Comparison of global electrochemical quantities of the resolved model (red) and the P2D model (blue) over time.}
\end{figure}

\begin{figure}[ht]
  \centering
  \begin{subfigure}[b]{0.45\textwidth}
    \centering
    \input{figures/comparison_resolved_homogenized_slice_concentration.tikz}
    \caption{Laterally (i.e., in the x-y-plane) averaged concentration at the end of discharge of the resolved model (red) and the P2D model (blue). For the dashed line, only the connected particles of the resolved model are considered.}
    \label{fig:comparison_resolved_homogenized_slice_concentration}
  \end{subfigure}
  \hfill
  \begin{subfigure}[b]{0.45\textwidth}
    \centering
    \def\svgwidth{0.9\textwidth}
\begingroup%
  \makeatletter%
  \providecommand\color[2][]{%
    \errmessage{(Inkscape) Color is used for the text in Inkscape, but the package 'color.sty' is not loaded}%
    \renewcommand\color[2][]{}%
  }%
  \providecommand\transparent[1]{%
    \errmessage{(Inkscape) Transparency is used (non-zero) for the text in Inkscape, but the package 'transparent.sty' is not loaded}%
    \renewcommand\transparent[1]{}%
  }%
  \providecommand\rotatebox[2]{#2}%
  \newcommand*\fsize{\dimexpr\f@size pt\relax}%
  \newcommand*\lineheight[1]{\fontsize{\fsize}{#1\fsize}\selectfont}%
  \ifx\svgwidth\undefined%
    \setlength{\unitlength}{687.33363009bp}%
    \ifx\svgscale\undefined%
      \relax%
    \else%
      \setlength{\unitlength}{\unitlength * \real{\svgscale}}%
    \fi%
  \else%
    \setlength{\unitlength}{\svgwidth}%
  \fi%
  \global\let\svgwidth\undefined%
  \global\let\svgscale\undefined%
  \makeatother%
  \begin{picture}(1,0.71038387)%
    \lineheight{1}%
    \setlength\tabcolsep{0pt}%
    \put(0,0){\includegraphics[width=\unitlength,page=1]{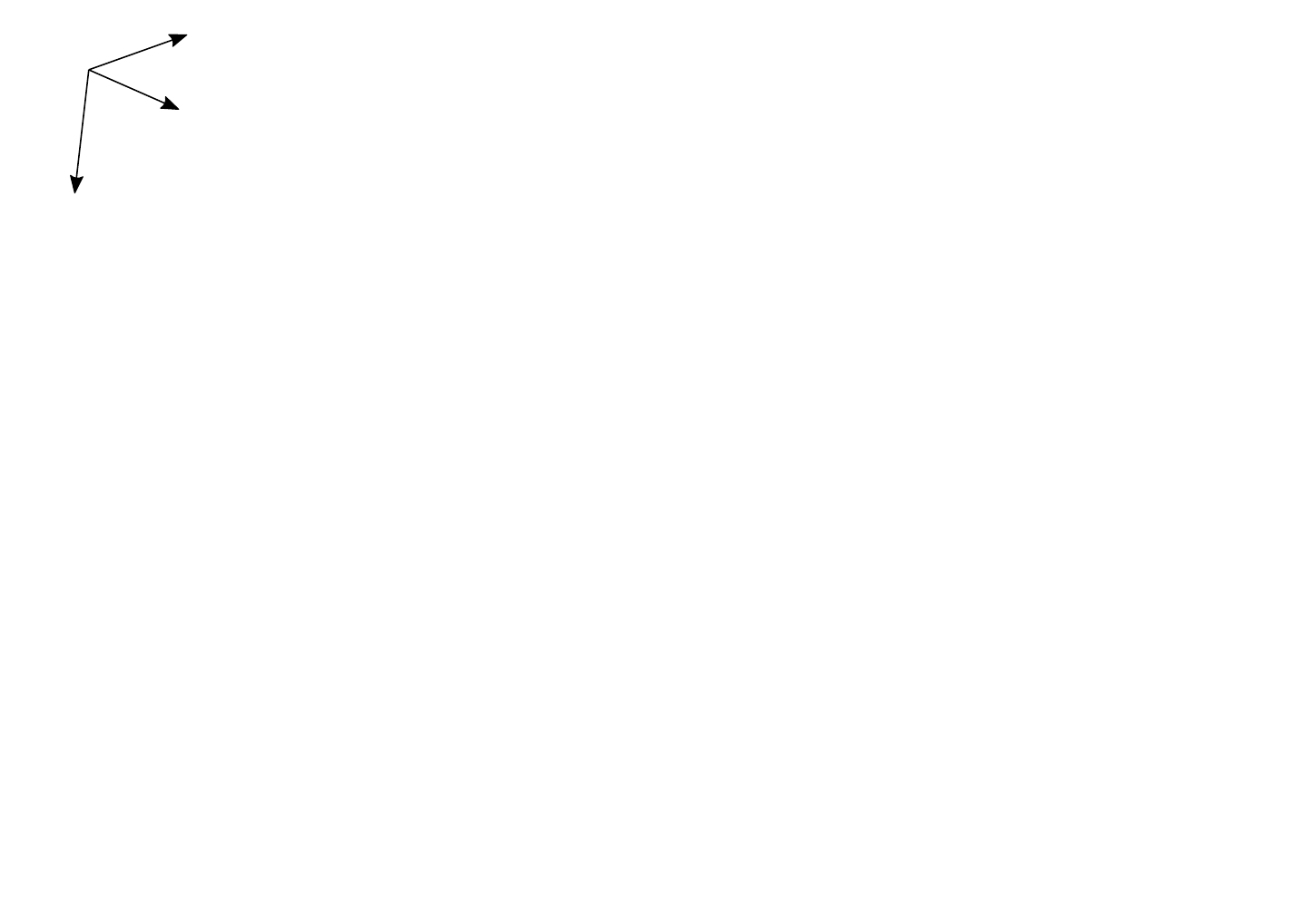}}%
    \put(-0.00096614,0.55028014){\color[rgb]{0,0,0}\makebox(0,0)[lt]{\lineheight{1.25}\smash{\begin{tabular}[t]{l}$z$\end{tabular}}}}%
    \put(0.1241233,0.70154082){\color[rgb]{0,0,0}\makebox(0,0)[lt]{\lineheight{1.25}\smash{\begin{tabular}[t]{l}$x$\end{tabular}}}}%
    \put(0.11996678,0.57695281){\color[rgb]{0,0,0}\makebox(0,0)[lt]{\lineheight{1.25}\smash{\begin{tabular}[t]{l}$y$\end{tabular}}}}%
    \put(0,0){\includegraphics[width=\unitlength,page=2]{figures/comparison_concentration_cathode_end_of_discharge.pdf}}%
  \end{picture}%
\endgroup%

    \caption{Concentration in the cathode at the end of discharge within the resolved model. The sandwich lithiation is visible.}
    \label{fig:comparison_concentration_cathode_end_of_discharge}
  \end{subfigure}
  \caption{Comparison of the concentration in the cathode at the end of discharge. Note that the end of discharge in this figure is determined by the resolved model.}
\end{figure}

\begin{figure}[ht]
  \centering
  \input{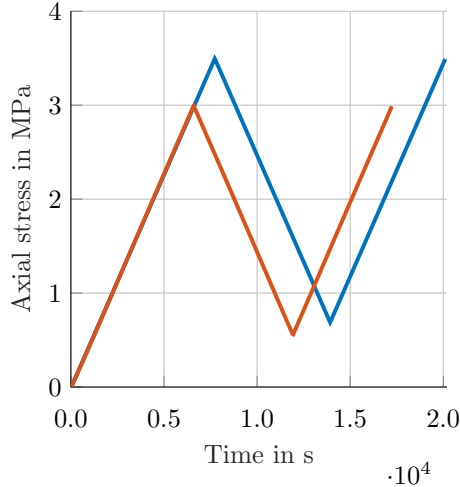}
  \caption{Comparison of axial stress computed with the resolved model (red) and the P2D model (blue).}
  \label{fig:comparison_resolved_homogenized_stress}
\end{figure}

\begin{figure}[ht]
  \centering
  \input{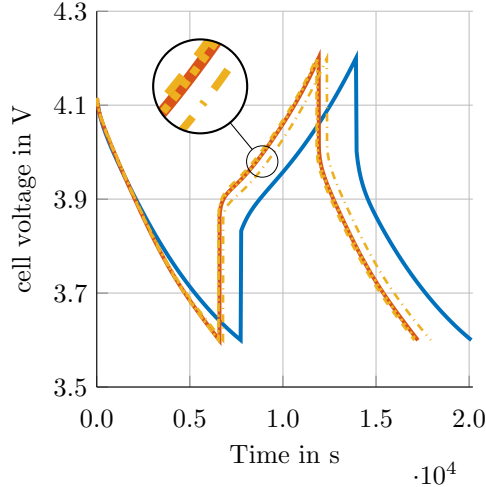}
  \caption{Analysis of the influence of different representations of the microstructure of the resolved model with different arrangements of the active material particles but with almost identical values of the porosity. The red line represents an evaluation of the resolved model, and the blue line of the P2D model. The three yellow lines denote the results of the resolved model with three different representations of the microstructure.}
  \label{fig:comparison_resolved_homogenized_resolved_variations}
\end{figure}

Global and locally resolved quantities computed with both models are compared for a discharge-charge-discharge sequence. The development of the cell voltage over time for the P2D as well as for the resolved model is compared in \cref{fig:comparison_resolved_homogenized_cell_voltage_c_rate}.
In the resolved case, the voltage limits are reached earlier. This leads to an overprediction of the amount of transported charge by the P2D model. The transported amount of charge is quantified in terms of the change in the state of charge (SOC). The SOC is defined as the normalized, integrated concentration in the cathode $\text{SOC}(t) = \frac{\int_{\Omega_\text{c}} c(t) - c_\text{min} \text{ d}\Omega}{\int_{\Omega_\text{c}} c_\text{max} - c_\text{min} \text{ d}\Omega}$. Its development over time is shown in \cref{fig:comparison_resolved_homogenized_soc}. One characteristic quantity is the amount of unreachable capacity, i.e., the SOC at the end of discharge. This value is indicated in \cref{fig:comparison_resolved_homogenized_soc} and has a relative deviation between both models of~$\epsilon = \left|\frac{\text{SOC}_\text{end,res} - \text{SOC}_\text{end,P2D}}{1-\text{SOC}_\text{end,res}}\right| \approx 13\% $. The deviation between both models might have two origins: (1) The reduced utilization in the resolved case and, therefore, a reduced available capacity and (2) the loss of geometric information of the complex microstructure in the P2D model that leads to a different total impedance. The influence of the as perfect assumed utilization in the P2D model is quantified by artificially reducing the available capacity within the P2D model by adjusting the prefactor~$A$ to~$A=0.320 \frac{\mu \text{m}^2}{\mu \text{m}^3}$. The result is shown in \cref{fig:comparison_resolved_homogenized_soc} by the additional dashed line. However, there is still a deviation between the SOC at the end of discharge computed with both models. Thus, the remaining deviation is attributed to a different intercalation behavior in the cathode due to the geometrically complex microstructure. The difference in the intercalation behavior is further studied by evaluating the laterally averaged concentration computed with the two models. The laterally averaged concentration is defined as
\begin{equation}
  \bar{c}_\text{avg}(z) = \frac{\int_x \int_y c(x,y,z) \text{d}x \text{d}y}{\int_x \int_y \text{d}x \text{d}y}.
\end{equation}
Note that the averaged concentration is smoothed by $c_\text{avg}(z) = \frac{1}{2 \epsilon} \int_{z-\epsilon}^{z+ \epsilon} \bar{c}_\text{avg}(\zeta) \text{ d} \zeta$ with the interval~$2 \epsilon = 10 \, \mu \text{m}$. The smoothed averaged concentration within the x-y-plane is shown in \cref{fig:comparison_resolved_homogenized_slice_concentration} over the axial coordinate for both models at the moment when the resolved model has reached the lower bound of the cell voltage.
For the resolved model, one evaluation of the smoothed averaged concentration considers all particles and another evaluation considers only the connected particles. A significant deviation between both models is visible. A sandwich lithiation, i.e., the favored lithiation close to the current collector and close to the anode, is observable within the connected particles of the resolved model as already reported in \cite{Neumann2020} with similar material parameters. In the P2D model, the sandwich lithiation is significantly less prominent but still observable. The spatially resolved concentration at the end of discharge computed with the resolved model is shown in \cref{fig:comparison_concentration_cathode_end_of_discharge}. There, the influence of the complex microstructure becomes visible: an increased concentration within small particles compared to larger particles, and the concentration in the non-connected particles remains at its initial value. This spatially more inhomogeneous concentration distribution originates from the complex conduction and diffusion paths within the resolved model. Especially, the diffusion within the differently sized particles is not sufficiently represented by the P2D model. The spatially different lithiation state leads to locally different equilibrium potentials and, thus, to a different development of the cell voltage between the models over time as observed in \cref{fig:comparison_resolved_homogenized_cell_voltage_c_rate}. \\
Finally, the averaged axial stress computed with both models is compared in \cref{fig:comparison_resolved_homogenized_stress}.
Again, the results of both models show a good agreement until the lower cut-off voltage is reached by the resolved model. The development of the axial stress in this setup is dominated by the volume change within the lithium metal anode, for which no homogenization is required and, thus, only minor differences are expected. This volume change depends linearly on the amount of transported charge. Due to the applied constant current, the transported amount in both models is identical until the lower cut-off voltage is reached, as discussed before.\\
Note that the results from the resolved model also depend on the artificial reduction of the lateral size of the SSB cell as shown in \cref{fig:3D_geometry_comparison}. For the assessment of this dependency, the cell voltage of multiple microstructures with different arrangements of the active material particles but with almost identical values of the porosity is evaluated and compared in \cref{fig:comparison_resolved_homogenized_resolved_variations}.
Only small deviations between the different results become visible, especially in comparison to the deviations between the resolved model and the P2D model. Therefore, it can be concluded that the chosen lateral size is sufficiently large.
\subsection{Comparison of the influence of crucial parameters}
The development of a novel or optimized design of an SSB cell requires the analysis of output quantities in a wide range of various input parameters and their combinations and not just one evaluation of the model at one working point, as done before. As one example, the influence of the variation of crucial input parameters in a wide range on the remaining capacity after the discharge is analyzed. The influence is quantified for individual input parameters and for a combination of several input parameters. As all relevant phenomena and interactions are already identifiable during the first discharge and the SOC at the end of discharge is almost identical after the first and the second discharge, we limit the analysis to only one discharge phase to reduce computational costs. Again, the remaining capacity is evaluated in terms of the SOC at the end of discharge. The electronic conductivity in the cathode, the diffusion coefficient in the cathode, the ionic conductivity in the solid electrolyte, and the applied discharge current are identified without an elaborate analysis as crucial parameters w.r.t. the SOC at the end of discharge. Their ranges, together with the statistical distributions, are listed in \cref{table:comparison_resolved_homogenized_intervals}.
\begin{table}[ht]
  \renewcommand{\arraystretch}{1.2}
  \centering
  \caption{Intervals of the crucial input parameters. \textit{Uniform in log-space} means that the logarithmic values of the samples are uniformly distributed.}
  \begin{tabular}{| c | c | c | c |}
    \hline
    \textbf{quantity}                             & \textbf{prob. distribution} & \textbf{min. value}                    & \textbf{max. value} \\
    \hline
    Electronic conductivity in cathode ($\sigma$) & uniform in log-space  & $10^{-2} \frac{\text{S}}{\text{m}}$    & $10 \frac{\text{S}}{\text{m}}$ \\
    Ionic conductivity in electrolyte ($\kappa$)  & uniform in log-space  & $10^{-3} \frac{\text{S}}{\text{m}}$    & $1 \frac{\text{S}}{\text{m}}$  \\
    Diffision coefficient in cathode ($D$)        & uniform in log-space  & $10^{-15} \frac{\text{m}^2}{\text{s}}$ & $10^{-12} \frac{\text{m}^2}{\text{s}}$ \\
    External current ($N_\rho$)                   & uniform               & $2.89 \frac{\text{A}}{\text{m}^2}$     & $9.64 \frac{\text{A}}{\text{m}^2}$ \\
    \hline
\end{tabular}
  \label{table:comparison_resolved_homogenized_intervals}
\end{table}
All parameters are independent of each other.
We begin with a comparison of the sensitivity of the input parameters based on the results of both models, followed by a physical interpretation of the deviations in the sensitivity.
\subsubsection{Comparison of the influence of all parameters}
\begin{figure}[ht]
  \centering
%
%
\definecolor{mycolor1}{rgb}{0.00000,0.44700,0.74100}%
\definecolor{mycolor2}{rgb}{0.92900,0.69400,0.12500}%
\begin{tikzpicture}

\begin{axis}[%
width=5.0cm,
height=5.0cm,
scale only axis,
xmin=0,
xmax=1,
x tick label style={
/pgf/number format/.cd,
fixed,
fixed zerofill,
precision=1,
/tikz/.cd,
yshift=-.5em},
xlabel style={font=\color{white!15!black}},
xlabel={${\text{SOC}_\text{end}}$ of resolved model},
ymin=0,
ymax=1,
y tick label style={
/pgf/number format/.cd,
fixed,
fixed zerofill,
precision=1,
/tikz/.cd},
ylabel style={font=\color{white!15!black}},
ylabel={${\text{SOC}_\text{end}}$ of P2D model},
axis background/.style={fill=white},
xmajorgrids,
ymajorgrids,
axis x line*=bottom,
axis y line*=left
]
\addplot [color=mycolor1, line width=2.0pt, only marks, mark size=4.0pt, mark=x, mark options={solid, mycolor1}, forget plot]
  table[row sep=crcr]{%
0.100537655474916	0.0376272734363899\\
0.704726688504343	0.726766048186825\\
0.102500641689398	0.0408955796654539\\
0.736455395663725	0.683474848423961\\
0.998500190579712	0.998907285834333\\
0.112146773261664	0.0429777246016509\\
0.258282488565362	0.310984313352469\\
0.182092933601206	0.0684584355018374\\
0.413924512567558	0.276748191550071\\
0.208608657619762	0.191115873897239\\
0.614378611987526	0.722507235052518\\
0.57492763058628	0.268514862879093\\
0.408098619367327	0.161522336165168\\
0.172586848112479	0.185385625470446\\
0.118634093725384	0.0473938245722768\\
0.556642882356345	0.58055060551467\\
0.12329609046905	0.0505806979293063\\
0.107753629399396	0.045173089659673\\
0.108200541685489	0.0449187598079749\\
0.951786414718943	0.922671529603051\\
0.602122108729442	0.474343478121686\\
0.14301069077927	0.0571858485483252\\
0.279151292306194	0.335444345693939\\
0.255906865083568	0.0494556399347103\\
0.816063883865379	0.897405420388531\\
0.183257683175827	0.0677489617361184\\
0.488850656866216	0.517964535161571\\
0.397923750223105	0.101575180788004\\
0.210183646795071	0.0411459603680331\\
0.973801352594487	0.977496696985767\\
0.11191640525075	0.0773727661198153\\
0.594897783519044	0.51032553403313\\
0.817544101384847	0.88858783733699\\
0.107798334684113	0.0397369266869169\\
0.247536456670892	0.274806530198447\\
0.296793426648708	0.0583763108730542\\
0.45833975162834	0.250694029899759\\
0.873869474189977	0.863659433367846\\
0.101712087387369	0.0390678026444299\\
0.49831901224523	0.436142373218288\\
0.394929926519392	0.200476102756028\\
0.506251097413361	0.503336031870074\\
0.106874867621883	0.0439387582916336\\
0.748093196749045	0.695921054564973\\
0.334486373211624	0.0575786839557024\\
0.119693529372374	0.0967025199445734\\
0.876978073383106	0.95775360700931\\
0.209018781175478	0.0781336433711243\\
0.238503737143558	0.316669976695793\\
0.209176317202422	0.080780814521494\\
0.669897945087239	0.73680955534593\\
0.47837179558647	0.230909794032127\\
0.606525690325411	0.427978347332067\\
0.192642646200109	0.215554867345216\\
0.104269010899495	0.0411635053255066\\
0.703190361124652	0.733615838529836\\
0.11364679954595	0.0409689745015236\\
0.926166536702806	0.949138796364716\\
0.100254993960337	0.0390302385986965\\
0.807334490018125	0.534936713873775\\
0.712724580468665	0.697455127973852\\
0.140865692016192	0.0511031803696975\\
0.557016548840495	0.578407459089934\\
0.13299487913089	0.0501075984998973\\
0.529884915005235	0.513296593859852\\
0.111411172197862	0.0430367558771843\\
0.570465383317037	0.663321222915428\\
0.283805887803585	0.0693442711299846\\
0.491926904540707	0.247414543747225\\
0.347249556269256	0.315640081228916\\
0.103549698036441	0.0405798386271608\\
0.938706771190072	0.941488648981463\\
0.286491613226197	0.0513445097280815\\
0.831900592017677	0.867200991754875\\
0.186712982876811	0.192204791692015\\
0.55775538401955	0.408492052164731\\
0.638039552456918	0.629007343556159\\
0.128980352239945	0.0512170531454107\\
0.233880635061945	0.303805573017403\\
0.151255635387663	0.0588292005458397\\
0.8787255975515	0.943721442494099\\
0.158251892160405	0.0611693709670484\\
0.140252873365034	0.165073480397361\\
0.551461454387668	0.375238489251089\\
0.524342728970988	0.25503317542564\\
0.563056643745091	0.619127291074847\\
0.11038018536005	0.0490888173163446\\
0.552826273354069	0.459337188166244\\
0.102179181667622	0.03797904132907\\
0.804467534977288	0.41099589866983\\
0.101129738860308	0.0390138180383351\\
0.949527849504943	0.834089275318415\\
0.527708087813853	0.409049720882686\\
0.194908573248583	0.0749800689507116\\
0.904553526189242	0.941055062047062\\
0.108161101821547	0.0402314599029273\\
0.110817721837933	0.0452395057489602\\
0.559180561273615	0.434755123727751\\
0.100146800422627	0.037568171068367\\
0.99647268140399	0.990272352213017\\
0.813063430198	0.705275592207782\\
0.116833052181639	0.0448467482282316\\
0.772591342460163	0.865263564276846\\
0.199015319291256	0.0569965407741167\\
0.534403248122535	0.53133383100508\\
0.11295967042417	0.0448842214587796\\
0.238553071800847	0.296207642633094\\
0.722796195084842	0.405213894283027\\
0.484932696048174	0.23204328945579\\
0.750412732439797	0.797327731294525\\
0.110114962273522	0.0491344734806872\\
0.109841573982162	0.0441308882003622\\
0.192668367000084	0.0484918987112486\\
0.567390629413105	0.587810034260071\\
0.107702911971406	0.0459339700000273\\
0.515158032373508	0.398224008064786\\
0.70741522773638	0.758573270376385\\
0.189724545319237	0.0704595124193232\\
0.111749865080993	0.0898042800411842\\
0.432410491552569	0.264883883897221\\
0.399671801254899	0.32872912567035\\
0.127958280387669	0.0498024417208229\\
0.946367437426702	0.972734567400394\\
0.52986313402558	0.217024966573648\\
0.32398374667839	0.0503575651749159\\
0.581403405058533	0.606878788121469\\
0.100403578639298	0.0381060222032142\\
0.875997502774358	0.870730032929182\\
0.10426197855285	0.0401419180000766\\
0.542290905433832	0.543338902892349\\
0.103779428464858	0.041599676256999\\
0.96972015265923	0.946530821115016\\
0.779851087761762	0.607553154117661\\
0.162290161121801	0.0672768158555503\\
0.943091827758502	0.963965908705799\\
0.105019006028894	0.0389758603018273\\
0.767109439376924	0.865658019378954\\
0.172114628924606	0.0650531101917375\\
0.101555831367716	0.0404550742628433\\
0.52379330597756	0.260950802133667\\
0.4518086231213	0.185476638550945\\
0.388512098513535	0.423724752339084\\
0.114448694024937	0.049086725462142\\
0.511133595671293	0.404202440614096\\
0.108525242356806	0.0423904900345284\\
0.633434456277592	0.638867058824776\\
0.136284907310497	0.144264718981664\\
0.473078194382569	0.315425363121072\\
0.755518243700132	0.689834789741404\\
0.144646438771531	0.0575665933875778\\
};
\addplot [color=mycolor2, line width=2.0pt, forget plot]
  table[row sep=crcr]{%
0	0\\
1	1\\
};
\end{axis}

\end{tikzpicture}%
  \caption{Comparison of the SOC at the end of discharge computed with the resolved model and the P2D model for 150 sets of parameters. A mark on the diagonal line indicates that the SOC at the end of discharge is equal, computed with both models for one set of parameters.}
  \label{fig:comparison_resolved_homogenized_diagonal}
\end{figure}
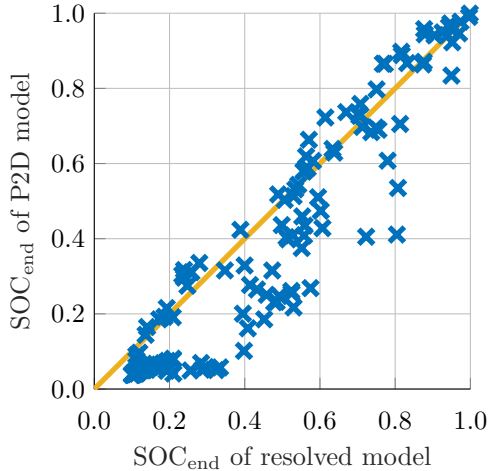

Both models are evaluated with the same 150 random samples. For one sample, a value is drawn for each of the parameters from the distributions listed in \cref{table:comparison_resolved_homogenized_intervals}. The SOC at the end of discharge is evaluated for each sample. Again, the cell is discharged from the fully charged state until the lower cell voltage of $\Delta \Phi_\text{low} = 3.6 \, \text{V}$ is reached. As a first visual indicator for a comparison between the two models, the result of~$\text{SOC}_\text{end}$ for one sample of input parameters evaluated with both models is compared in \cref{fig:comparison_resolved_homogenized_diagonal}. There, one mark indicates one sample of the input parameters, i.e., one combination of the four parameters.
The position of the marks within the diagram is determined by the SOC at the end of discharge: The results from the resolved model are aligned on the horizontal axis, and the results from the P2D model on the vertical axis. Thus, marks on the diagonal indicate the same results computed with both models, marks below the diagonal indicate that the P2D model underestimates the remaining capacity, and marks above the diagonal indicate that the P2D model overestimates the remaining capacity. \\
At a first sight, it can be observed that the selected parameters have a significant influence as the SOC at the end of discharge varies between 1 and almost 0. Furthermore, it can be concluded that both models express a similar trend as the marks become scarce far away from the diagonal line with no marks at the top-left and low-right corners. The majority of the marks are below the main diagonal. This means that the P2D model overpredicts the amount of transported charge. Especially for small values of the SOC at the end of discharge, the deviation between both models is up to factor five.
\subsubsection{Comparison of global sensitivites}
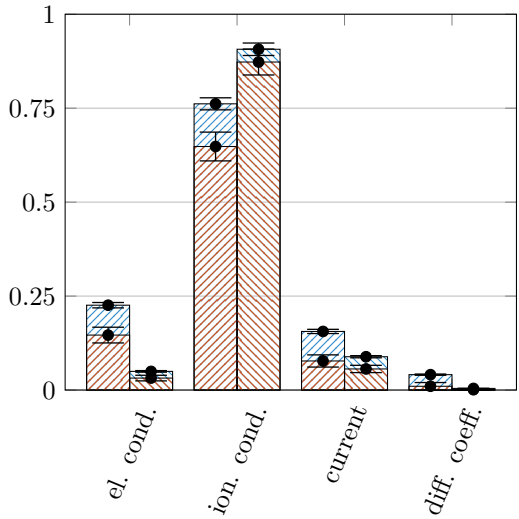
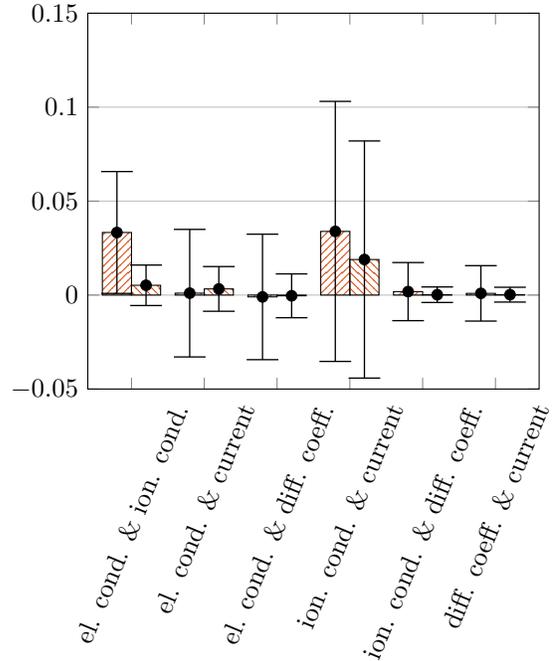
\begin{figure}[ht]
  \centering
  \begin{subfigure}[b]{0.45\textwidth}
    \centering
%
%
\definecolor{mycolor1}{rgb}{0.00000,0.44706,0.74118}%
\definecolor{mycolor2}{rgb}{0.85098,0.32549,0.09804}%
\begin{tikzpicture}

\begin{axis}[%
width=6.0cm,
height=5.0cm,
scale only axis,
xmin=0.4,
xmax=4.6,
xtick={1,2,3,4},
xticklabels={{el. cond.},{ion. cond.},{current},{diff. coeff.}},
xticklabel style={rotate=67.5},
ytick={0,0.25,0.5,0.75,1.0},
ymin=0,
ymax=1.0,
y tick label style={
/pgf/number format/.cd,
/tikz/.cd},
ymajorgrids,
axis background/.style={fill=white}
]
\addplot[ybar, bar width=0.4, draw=black, area legend, pattern=north east lines, pattern color=mycolor1!80] table[row sep=crcr] {%
0.8	0.225760427696638\\
1.8	0.761590725709228\\
2.8	0.155950768831105\\
3.8	0.0409368747532135\\
};
\addplot[ybar, bar width=0.4, draw=black, area legend, pattern=north west lines, pattern color=mycolor1!80] table[row sep=crcr] {%
1.2	0.049643032001919\\
2.2	0.906800278640025\\
3.2	0.0886305313095527\\
4.2	0.00429575436634175\\
};
\addplot[ybar, bar width=0.4, draw=black, area legend, pattern=north east lines, pattern color=mycolor2] table[row sep=crcr] {%
0.8	0.146243969526853\\
1.8	0.648020081550134\\
2.8	0.0773451553194495\\
3.8	0.0100151285069117\\
};
\addplot[ybar, bar width=0.4, draw=black, area legend, pattern=north west lines, pattern color=mycolor2] table[row sep=crcr] {%
1.2	0.0318366005236328\\
2.2	0.872840815798757\\
3.2	0.0560936945870537\\
4.2	0.000797779047605094\\
};
\addplot [color=black, only marks, forget plot]
 plot [error bars/.cd, y dir=both, y explicit, error bar style={line width=0.5pt}, error mark options={line width=0.5pt, mark size=6.0pt, rotate=90}]
 table[row sep=crcr, y error plus index=2, y error minus index=3]{%
0.8	0.225760427696638	0.0071471989566404	0.0071471989566404\\
1.8	0.761590725709228	0.0160401513669798	0.0160401513669798\\
2.8	0.155950768831105	0.00559438065497068	0.00559438065497068\\
3.8	0.0409368747532135	0.00161238792419575	0.00161238792419575\\
};
\addplot [color=black, only marks, forget plot]
 plot [error bars/.cd, y dir=both, y explicit, error bar style={line width=0.5pt}, error mark options={line width=0.5pt, mark size=6.0pt, rotate=90}]
 table[row sep=crcr, y error plus index=2, y error minus index=3]{%
1.2	0.049643032001919	0.0020871486781985	0.0020871486781985\\
2.2	0.906800278640025	0.0165893180450425	0.0165893180450425\\
3.2	0.0886305313095527	0.00287562899373159	0.00287562899373159\\
4.2	0.00429575436634175	0.000208200456810041	0.000208200456810041\\
};
\addplot [color=black, only marks, forget plot]
 plot [error bars/.cd, y dir=both, y explicit, error bar style={line width=0.5pt}, error mark options={line width=0.5pt, mark size=6.0pt, rotate=90}]
 table[row sep=crcr, y error plus index=2, y error minus index=3]{%
0.8	0.146243969526853	0.0209276716821425	0.0209276716821425\\
1.8	0.648020081550134	0.0383041333844994	0.0383041333844994\\
2.8	0.0773451553194495	0.0162015370962289	0.0162015370962289\\
3.8	0.0100151285069117	0.00994615564257976	0.00994615564257976\\
};
\addplot [color=black, only marks, forget plot]
 plot [error bars/.cd, y dir=both, y explicit, error bar style={line width=0.5pt}, error mark options={line width=0.5pt, mark size=6.0pt, rotate=90}]
 table[row sep=crcr, y error plus index=2, y error minus index=3]{%
1.2	0.0318366005236328	0.00748593170805279	0.00748593170805279\\
2.2	0.872840815798757	0.0344826656360585	0.0344826656360585\\
3.2	0.0560936945870537	0.00967811728689543	0.00967811728689543\\
4.2	0.000797779047605094	0.00265533126799021	0.00265533126799021\\
};
\end{axis}
\end{tikzpicture}%
    \caption{The total-order (blue) and first-order Sobol indices (red).}
    \label{fig:comparison_resolved_homogenized_sobol_indices_first_order_merged}
  \end{subfigure}
  \hfill
  \begin{subfigure}[b]{0.45\textwidth}
    \centering
%
%
\definecolor{mycolor1}{rgb}{0.85000,0.32500,0.09800}%
\begin{tikzpicture}

\begin{axis}[%
width=6.0cm,
height=5.0cm,
scale only axis,
xmin=0.4,
xmax=6.6,
xtick={1,2,3,4,5,6},
xticklabels={{el. cond. \& ion. cond.},{el. cond. \& current},{el. cond. \& diff. coeff.},{ion. cond. \& current},{ion. cond. \& diff. coeff.},{diff. coeff. \& current}},
xticklabel style={rotate=67.5},
ymin=-0.05,
ymax=0.15,
ytick={-0.05,0.0,0.05,0.1,0.15},
y tick label style={
/pgf/number format/.cd,
fixed,
precision=2,
/tikz/.cd},
ymajorgrids,
axis background/.style={fill=white}
]
\addplot[ybar, bar width=0.4, draw=black, area legend, pattern=north east lines, pattern color=mycolor1] table[row sep=crcr] {%
0.8	0.0333063399355894\\
1.8	0.00101698607586462\\
2.8	-0.000982409296986552\\
3.8	0.0338897203706772\\
4.8	0.00184925174435373\\
5.8	0.000909406589335622\\
};
\addplot[ybar, bar width=0.4, draw=black, area legend, pattern=north west lines, pattern color=mycolor1] table[row sep=crcr] {%
1.2	0.00520686671994249\\
2.2	0.00327723248957464\\
3.2	-0.000387835667256086\\
4.2	0.0189166494164463\\
5.2	0.000196087366486548\\
6.2	0.000220572748460541\\
};
\addplot [color=black, only marks, forget plot]
 plot [error bars/.cd, y dir=both, y explicit, error bar style={line width=0.5pt}, error mark options={line width=0.5pt, mark size=6.0pt, rotate=90}]
 table[row sep=crcr, y error plus index=2, y error minus index=3]{%
0.8	0.0333063399355894	0.032428256663788	0.032428256663788\\
1.8	0.00101698607586462	0.0339454503436206	0.0339454503436206\\
2.8	-0.000982409296986552	0.0333971339470389	0.0333971339470389\\
3.8	0.0338897203706772	0.0691959116428466	0.0691959116428466\\
4.8	0.00184925174435373	0.0154626421333318	0.0154626421333318\\
5.8	0.000909406589335622	0.0147457013630147	0.0147457013630147\\
};
\addplot [color=black, only marks, forget plot]
 plot [error bars/.cd, y dir=both, y explicit, error bar style={line width=0.5pt}, error mark options={line width=0.5pt, mark size=6.0pt, rotate=90}]
 table[row sep=crcr, y error plus index=2, y error minus index=3]{%
1.2	0.00520686671994249	0.0107728690635492	0.0107728690635492\\
2.2	0.00327723248957464	0.0118938865423198	0.0118938865423198\\
3.2	-0.000387835667256086	0.0116615386166425	0.0116615386166425\\
4.2	0.0189166494164463	0.0630992078520958	0.0630992078520958\\
5.2	0.000196087366486548	0.00413419269458936	0.00413419269458936\\
6.2	0.000220572748460541	0.00394195429043748	0.00394195429043748\\
};
\end{axis}
\end{tikzpicture}%
    \caption{The second-order Sobol indices.}
    \label{fig:comparison_resolved_homogenized_sobol_indices_second_order_merged}
  \end{subfigure}
  \caption{The Sobol indices including a~$95\%$ confidence interval. The bars on the left-hand side of each column indicate the computation based on the resolved model, and the bars on the right-hand side indicate the computation based on the P2D model.}
\end{figure}

A further analysis of the origin of the deviations in \cref{fig:comparison_resolved_homogenized_diagonal} is performed by comparing the sensitivity of the SOC at the end of discharge on the input parameters in terms of Sobol indices (refer to Appendix \ref{sec:sobol_indices} for a summary on the computation of the Sobol indices). It is important to note that the sensitivity analysis is in a global sense, i.e., not for one parameter set. Additionally, it is not restricted to individual parameters but also considers interactions between them by higher order Sobol indices. For the input parameters~$\Psi$ with a \textit{uniform distribution in log-space}, their logarithmic value with base 10 is used to compute the Sobol index, i.e., $\text{lg}(\frac{\Psi}{\Psi_1})$, where~$\Psi_1$ equals~$1 \frac{\text{S}}{\text{m}}$ in the case of the electronic and ionic conductivity and~$1 \frac{\text{m}^2}{\text{s}}$ in the case of the diffusion coefficient. A large number of model evaluations is required to obtain Sobol indices with a sufficient confidence interval due to the underlying Monte Carlo integration. Hence, a nonlinear meta-model based on a Gaussian process is trained with the 150 training samples from the previous example to circumvent the need for a large number of expensive model evaluations and thereby reduce the computational costs for the sensitivity analysis. Then, the Monte Carlo integration is performed using this significantly faster meta-model by drawing $2^{14} = 16 \, 384$ Monte Carlo samples from the input parameters and calculating the SOC at the end of discharge by evaluations of the meta-model. In~\cite{Wirthl2023}, this approach has been demonstrated to be very useful for a sensitivity analysis in complex systems. Based on this, the Sobol indices are computed. The Sobol indices of first and total order are shown in \cref{fig:comparison_resolved_homogenized_sobol_indices_first_order_merged} for both models.
The Sobol indices of total order quantify if a parameter has an influence at all. The Sobol indices of first order quantify if one parameter has an influence independent of the value of the other parameters. Similarities become visible: the ionic conductivity has the greatest influence within both models. Furthermore, the first-order index has almost the same value as the total-order index for all parameters, meaning that the combined influence of parameters is small. \\
However, differences occur as well: the difference between the first and total order indices computed with the P2D model is smaller for all parameters. This means that the P2D model underestimates the interactions between different physical phenomena. Furthermore, the P2D model predicts almost no influence of the diffusion coefficient, while the resolved model assigns a non-negligible influence. The P2D model underestimates the influence of the electronic conductivity within the composite cathode. Consequently, the Sobol index of first-order assigned to the ionic conductivity is smaller within the resolved model as all indices have to sum up to one. The second-order indices are shown in \cref{fig:comparison_resolved_homogenized_sobol_indices_second_order_merged}. From the second-order index, i.e., the combined influence of two parameters, it becomes obvious that the P2D model underestimates the interaction between physical phenomena, as the second-order Sobol-indices computed with the P2D model are smaller for all parameter combinations. \\
The Sobol indices of first and second order sum almost up to the total order indices, such that the computationally expensive evaluation of the Sobol indices of third order, i.e., the combined influence of three parameters, and fourth order, i.e., the combined influence of all parameters, is not required anymore.
\subsubsection{Comparison of the effects of single parameter variations}
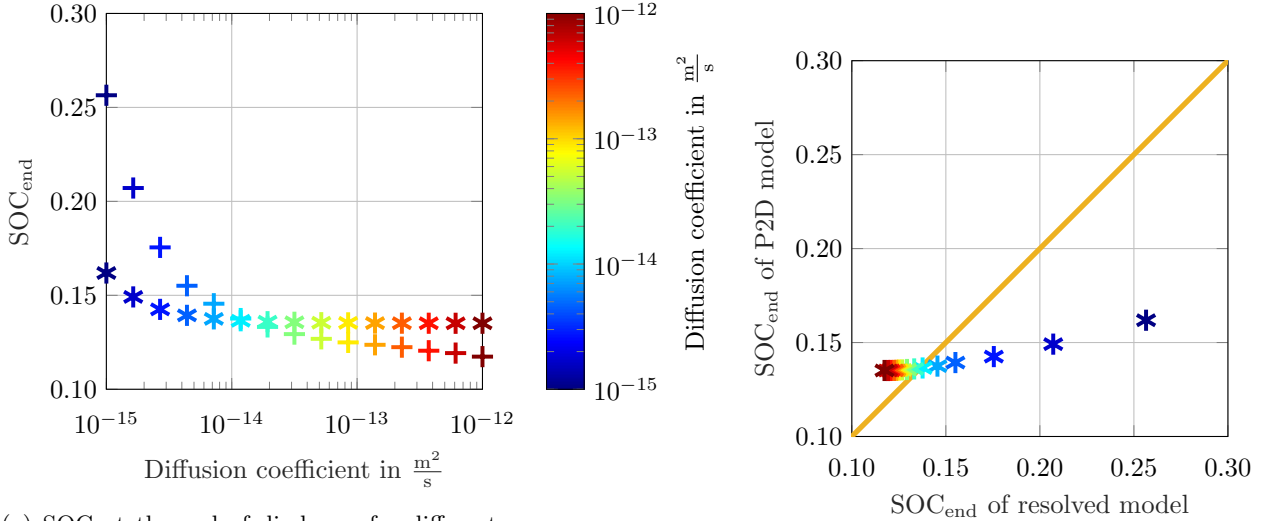
\begin{figure}[ht]
  \centering
  \begin{subfigure}[b]{0.4\textwidth}
    \centering
%
%
\definecolor{mycolor1}{rgb}{0.00000,0.00000,0.51562}%
\definecolor{mycolor2}{rgb}{0.00000,0.00000,0.79688}%
\definecolor{mycolor3}{rgb}{0.00000,0.07812,1.00000}%
\definecolor{mycolor4}{rgb}{0.00000,0.37500,1.00000}%
\definecolor{mycolor5}{rgb}{0.00000,0.65625,1.00000}%
\definecolor{mycolor6}{rgb}{0.00000,0.93750,1.00000}%
\definecolor{mycolor7}{rgb}{0.21875,1.00000,0.78125}%
\definecolor{mycolor8}{rgb}{0.51562,1.00000,0.48438}%
\definecolor{mycolor9}{rgb}{0.79688,1.00000,0.20312}%
\definecolor{mycolor10}{rgb}{1.00000,0.92188,0.00000}%
\definecolor{mycolor11}{rgb}{1.00000,0.64062,0.00000}%
\definecolor{mycolor12}{rgb}{1.00000,0.35938,0.00000}%
\definecolor{mycolor13}{rgb}{1.00000,0.06250,0.00000}%
\definecolor{mycolor14}{rgb}{0.78125,0.00000,0.00000}%
\definecolor{mycolor15}{rgb}{0.50000,0.00000,0.00000}%
\definecolor{mycolor16}{rgb}{0.92900,0.69400,0.12500}%
\begin{tikzpicture}

\begin{axis}[%
width=5.0cm,
height=5.0cm,
point meta min=1e-15,
point meta max=1e-12,
scale only axis,
xmode=log,
xmin=1e-15,
xmax=1e-12,
x tick label style={
/pgf/number format/.cd,
precision=0,
/tikz/.cd,
yshift=-.5em},
xlabel style={font=\color{white!15!black}},
xlabel={Diffusion coefficient in $\frac{\text{m}^2}{\text{s}}$},
ymin=0.1,
ymax=0.3,
y tick label style={
/pgf/number format/.cd,
fixed,
fixed zerofill,
precision=2,
/tikz/.cd},
ylabel style={font=\color{white!15!black}},
ylabel={$\text{SOC}_\text{end}$},
axis background/.style={fill=white},
xmajorgrids,
ymajorgrids,
colorbar,
colorbar style={
    ylabel=Diffusion coefficient in $\frac{\text{m}^2}{\text{s}}$,
    ymode=log,
    yticklabel style={
        /pgf/number format/.cd,
        fixed,
        precision=0,
    }
},
colormap/jet
]
\addplot [color=mycolor1, line width=1.5pt, only marks, mark size=4.0pt, mark=asterisk, mark options={solid, mycolor1}, forget plot]
  table[row sep=crcr]{%
1e-15	0.161874725035417\\
};
\addplot [color=mycolor1, line width=1.5pt, only marks, mark size=4.0pt, mark=+, mark options={solid, mycolor1}, forget plot]
  table[row sep=crcr]{%
1e-15	0.256468800649772\\
};
\addplot [color=mycolor2, line width=1.5pt, only marks, mark size=4.0pt, mark=asterisk, mark options={solid, mycolor2}, forget plot]
  table[row sep=crcr]{%
1.63789370695406e-15	0.149149505172117\\
};
\addplot [color=mycolor2, line width=1.5pt, only marks, mark size=4.0pt, mark=+, mark options={solid, mycolor2}, forget plot]
  table[row sep=crcr]{%
1.63789370695406e-15	0.207105979146006\\
};
\addplot [color=mycolor3, line width=1.5pt, only marks, mark size=4.0pt, mark=asterisk, mark options={solid, mycolor3}, forget plot]
  table[row sep=crcr]{%
2.68269579527973e-15	0.142518799140895\\
};
\addplot [color=mycolor3, line width=1.5pt, only marks, mark size=4.0pt, mark=+, mark options={solid, mycolor3}, forget plot]
  table[row sep=crcr]{%
2.68269579527973e-15	0.175510068525839\\
};
\addplot [color=mycolor4, line width=1.5pt, only marks, mark size=4.0pt, mark=asterisk, mark options={solid, mycolor4}, forget plot]
  table[row sep=crcr]{%
4.39397056076079e-15	0.139339740595734\\
};
\addplot [color=mycolor4, line width=1.5pt, only marks, mark size=4.0pt, mark=+, mark options={solid, mycolor4}, forget plot]
  table[row sep=crcr]{%
4.39397056076079e-15	0.155085232289991\\
};
\addplot [color=mycolor5, line width=1.5pt, only marks, mark size=4.0pt, mark=asterisk, mark options={solid, mycolor5}, forget plot]
  table[row sep=crcr]{%
7.19685673001153e-15	0.137463537228467\\
};
\addplot [color=mycolor5, line width=1.5pt, only marks, mark size=4.0pt, mark=+, mark options={solid, mycolor5}, forget plot]
  table[row sep=crcr]{%
7.19685673001153e-15	0.145529458641987\\
};
\addplot [color=mycolor6, line width=1.5pt, only marks, mark size=4.0pt, mark=asterisk, mark options={solid, mycolor6}, forget plot]
  table[row sep=crcr]{%
1.17876863479359e-14	0.136461987822743\\
};
\addplot [color=mycolor6, line width=1.5pt, only marks, mark size=4.0pt, mark=+, mark options={solid, mycolor6}, forget plot]
  table[row sep=crcr]{%
1.17876863479359e-14	0.13764023285198\\
};
\addplot [color=mycolor7, line width=1.5pt, only marks, mark size=4.0pt, mark=asterisk, mark options={solid, mycolor7}, forget plot]
  table[row sep=crcr]{%
1.93069772888325e-14	0.135958757678249\\
};
\addplot [color=mycolor7, line width=1.5pt, only marks, mark size=4.0pt, mark=+, mark options={solid, mycolor7}, forget plot]
  table[row sep=crcr]{%
1.93069772888325e-14	0.133170141702686\\
};
\addplot [color=mycolor8, line width=1.5pt, only marks, mark size=4.0pt, mark=asterisk, mark options={solid, mycolor8}, forget plot]
  table[row sep=crcr]{%
3.16227766016838e-14	0.135655145693413\\
};
\addplot [color=mycolor8, line width=1.5pt, only marks, mark size=4.0pt, mark=+, mark options={solid, mycolor8}, forget plot]
  table[row sep=crcr]{%
3.16227766016838e-14	0.129353411539656\\
};
\addplot [color=mycolor9, line width=1.5pt, only marks, mark size=4.0pt, mark=asterisk, mark options={solid, mycolor9}, forget plot]
  table[row sep=crcr]{%
5.1794746792312e-14	0.135427568421294\\
};
\addplot [color=mycolor9, line width=1.5pt, only marks, mark size=4.0pt, mark=+, mark options={solid, mycolor9}, forget plot]
  table[row sep=crcr]{%
5.1794746792312e-14	0.126834644825478\\
};
\addplot [color=mycolor10, line width=1.5pt, only marks, mark size=4.0pt, mark=asterisk, mark options={solid, mycolor10}, forget plot]
  table[row sep=crcr]{%
8.48342898244073e-14	0.13535318553849\\
};
\addplot [color=mycolor10, line width=1.5pt, only marks, mark size=4.0pt, mark=+, mark options={solid, mycolor10}, forget plot]
  table[row sep=crcr]{%
8.48342898244073e-14	0.124942064443615\\
};
\addplot [color=mycolor11, line width=1.5pt, only marks, mark size=4.0pt, mark=asterisk, mark options={solid, mycolor11}, forget plot]
  table[row sep=crcr]{%
1.38949549437314e-13	0.135222464674951\\
};
\addplot [color=mycolor11, line width=1.5pt, only marks, mark size=4.0pt, mark=+, mark options={solid, mycolor11}, forget plot]
  table[row sep=crcr]{%
1.38949549437314e-13	0.123662572197438\\
};
\addplot [color=mycolor12, line width=1.5pt, only marks, mark size=4.0pt, mark=asterisk, mark options={solid, mycolor12}, forget plot]
  table[row sep=crcr]{%
2.27584592607479e-13	0.135171976604695\\
};
\addplot [color=mycolor12, line width=1.5pt, only marks, mark size=4.0pt, mark=+, mark options={solid, mycolor12}, forget plot]
  table[row sep=crcr]{%
2.27584592607479e-13	0.122404625754894\\
};
\addplot [color=mycolor13, line width=1.5pt, only marks, mark size=4.0pt, mark=asterisk, mark options={solid, mycolor13}, forget plot]
  table[row sep=crcr]{%
3.72759372031494e-13	0.135143322826536\\
};
\addplot [color=mycolor13, line width=1.5pt, only marks, mark size=4.0pt, mark=+, mark options={solid, mycolor13}, forget plot]
  table[row sep=crcr]{%
3.72759372031494e-13	0.120514948872386\\
};
\addplot [color=mycolor14, line width=1.5pt, only marks, mark size=4.0pt, mark=asterisk, mark options={solid, mycolor14}, forget plot]
  table[row sep=crcr]{%
6.10540229658531e-13	0.135117247362697\\
};
\addplot [color=mycolor14, line width=1.5pt, only marks, mark size=4.0pt, mark=+, mark options={solid, mycolor14}, forget plot]
  table[row sep=crcr]{%
6.10540229658531e-13	0.119253108400915\\
};
\addplot [color=mycolor15, line width=1.5pt, only marks, mark size=4.0pt, mark=asterisk, mark options={solid, mycolor15}, forget plot]
  table[row sep=crcr]{%
1e-12	0.13512902878063\\
};
\addplot [color=mycolor15, line width=1.5pt, only marks, mark size=4.0pt, mark=+, mark options={solid, mycolor15}, forget plot]
  table[row sep=crcr]{%
1e-12	0.117366603643639\\
};
\end{axis}

\end{tikzpicture}%
    \caption{SOC at the end of discharge for different values of the diffusion coefficient with the resolved (+) and the P2D model ($\ast$).}
    \label{fig:comparison_resolved_homogenized_diffusion_result}
  \end{subfigure}
  \hfill
  \begin{subfigure}[b]{0.4\textwidth}
    \centering
%
%
\definecolor{mycolor1}{rgb}{0.00000,0.00000,0.51562}%
\definecolor{mycolor2}{rgb}{0.00000,0.00000,0.79688}%
\definecolor{mycolor3}{rgb}{0.00000,0.07812,1.00000}%
\definecolor{mycolor4}{rgb}{0.00000,0.37500,1.00000}%
\definecolor{mycolor5}{rgb}{0.00000,0.65625,1.00000}%
\definecolor{mycolor6}{rgb}{0.00000,0.93750,1.00000}%
\definecolor{mycolor7}{rgb}{0.21875,1.00000,0.78125}%
\definecolor{mycolor8}{rgb}{0.51562,1.00000,0.48438}%
\definecolor{mycolor9}{rgb}{0.79688,1.00000,0.20312}%
\definecolor{mycolor10}{rgb}{1.00000,0.92188,0.00000}%
\definecolor{mycolor11}{rgb}{1.00000,0.64062,0.00000}%
\definecolor{mycolor12}{rgb}{1.00000,0.35938,0.00000}%
\definecolor{mycolor13}{rgb}{1.00000,0.06250,0.00000}%
\definecolor{mycolor14}{rgb}{0.78125,0.00000,0.00000}%
\definecolor{mycolor15}{rgb}{0.50000,0.00000,0.00000}%
\definecolor{mycolor16}{rgb}{0.92900,0.69400,0.12500}%
\begin{tikzpicture}

\begin{axis}[%
width=5.0cm,
height=5.0cm,
point meta min=-15.0,
point meta max=-12.0,
scale only axis,
xmin=0.1,
xmax=0.3,
x tick label style={
/pgf/number format/.cd,
fixed,
fixed zerofill,
precision=2,
/tikz/.cd,
yshift=-.5em},
xlabel style={font=\color{white!15!black}},
xlabel={$\text{SOC}_\text{end}$ of resolved model},
ymin=0.1,
ymax=0.3,
y tick label style={
/pgf/number format/.cd,
fixed,
fixed zerofill,
precision=2,
/tikz/.cd},
ylabel style={font=\color{white!15!black}},
ylabel={$\text{SOC}_\text{end}$ of P2D model},
axis background/.style={fill=white},
xmajorgrids,
ymajorgrids
]
\addplot [color=mycolor1, line width=1.5pt, only marks, mark size=4.0pt, mark=asterisk, mark options={solid, mycolor1}, forget plot]
  table[row sep=crcr]{%
0.256468800649772	0.161874725035417\\
};
\addplot [color=mycolor2, line width=1.5pt, only marks, mark size=4.0pt, mark=asterisk, mark options={solid, mycolor2}, forget plot]
  table[row sep=crcr]{%
0.207105979146006	0.149149505172117\\
};
\addplot [color=mycolor3, line width=1.5pt, only marks, mark size=4.0pt, mark=asterisk, mark options={solid, mycolor3}, forget plot]
  table[row sep=crcr]{%
0.175510068525839	0.142518799140895\\
};
\addplot [color=mycolor4, line width=1.5pt, only marks, mark size=4.0pt, mark=asterisk, mark options={solid, mycolor4}, forget plot]
  table[row sep=crcr]{%
0.155085232289991	0.139339740595734\\
};
\addplot [color=mycolor5, line width=1.5pt, only marks, mark size=4.0pt, mark=asterisk, mark options={solid, mycolor5}, forget plot]
  table[row sep=crcr]{%
0.145529458641987	0.137463537228467\\
};
\addplot [color=mycolor6, line width=1.5pt, only marks, mark size=4.0pt, mark=asterisk, mark options={solid, mycolor6}, forget plot]
  table[row sep=crcr]{%
0.13764023285198	0.136461987822743\\
};
\addplot [color=mycolor7, line width=1.5pt, only marks, mark size=4.0pt, mark=asterisk, mark options={solid, mycolor7}, forget plot]
  table[row sep=crcr]{%
0.133170141702686	0.135958757678249\\
};
\addplot [color=mycolor8, line width=1.5pt, only marks, mark size=4.0pt, mark=asterisk, mark options={solid, mycolor8}, forget plot]
  table[row sep=crcr]{%
0.129353411539656	0.135655145693413\\
};
\addplot [color=mycolor9, line width=1.5pt, only marks, mark size=4.0pt, mark=asterisk, mark options={solid, mycolor9}, forget plot]
  table[row sep=crcr]{%
0.126834644825478	0.135427568421294\\
};
\addplot [color=mycolor10, line width=1.5pt, only marks, mark size=4.0pt, mark=asterisk, mark options={solid, mycolor10}, forget plot]
  table[row sep=crcr]{%
0.124942064443615	0.13535318553849\\
};
\addplot [color=mycolor11, line width=1.5pt, only marks, mark size=4.0pt, mark=asterisk, mark options={solid, mycolor11}, forget plot]
  table[row sep=crcr]{%
0.123662572197438	0.135222464674951\\
};
\addplot [color=mycolor12, line width=1.5pt, only marks, mark size=4.0pt, mark=asterisk, mark options={solid, mycolor12}, forget plot]
  table[row sep=crcr]{%
0.122404625754894	0.135171976604695\\
};
\addplot [color=mycolor13, line width=1.5pt, only marks, mark size=4.0pt, mark=asterisk, mark options={solid, mycolor13}, forget plot]
  table[row sep=crcr]{%
0.120514948872386	0.135143322826536\\
};
\addplot [color=mycolor14, line width=1.5pt, only marks, mark size=4.0pt, mark=asterisk, mark options={solid, mycolor14}, forget plot]
  table[row sep=crcr]{%
0.119253108400915	0.135117247362697\\
};
\addplot [color=mycolor15, line width=1.5pt, only marks, mark size=4.0pt, mark=asterisk, mark options={solid, mycolor15}, forget plot]
  table[row sep=crcr]{%
0.117366603643639	0.13512902878063\\
};
\addplot [color=mycolor16, line width=2.0pt, forget plot]
  table[row sep=crcr]{%
0	0\\
1	1\\
};
\end{axis}

\end{tikzpicture}%
    \caption{Comparison of the SOC at the end of discharge computed with both models.}
    \label{fig:comparison_resolved_homogenized_diffusion_result_diagonal}
  \end{subfigure}
  \caption{Comparison of the SOC at the end of discharge for different values of the diffusion coefficient.}
\end{figure}

Further insights into and a physical interpretation of the differences between the models can be gained by keeping all except for one parameter fixed. Here, the most interesting parameters are the diffusion coefficient, as the P2D model predicts almost no influence of it while the resolved model assigns an influence, and the ionic conductivity, which is the most influential parameter.\\
For this investigation, the diffusion coefficient is varied within the bounds given in \cref{table:comparison_resolved_homogenized_intervals} while the electronic conductivity $\sigma = 0.32 \frac{\text{S}}{\text{m}}$, the ionic conductivity $\kappa = 0.032 \frac{\text{S}}{\text{m}}$, and the current $N_\rho = 6.27 \frac{\text{A}}{\text{m}^2}$ are fixed to their mean values. In \cref{fig:comparison_resolved_homogenized_diffusion_result}, the SOC at the end of discharge computed with both models is shown for these parameters.
The results from both models follow a similar trend. The SOC at the end of discharge approaches a progressive increase for small values of the diffusion coefficient and a regressive decrease for large values. However, two deviations become visible: First, the result of the P2D model approaches a constant value for large values of the diffusion coefficient while the result of the resolved model continues to decrease. The reason that the P2D model approaches a constant value is attributed to the non-connected particles as a central modeling assumption of the P2D model (c.f. \cref{eq:homogenized_conservation_mass,eq:homogenized_coupling}). Thus, diffusion between particles is not possible, such that a further increase of the diffusion coefficient does not lead to more transportation by diffusion through the cathode instead of conduction through the solid electrolyte from a certain point on. In this case, other transport phenomena like conduction are the limiting factors. In the resolved model, the large diffusion results in inter-particle diffusion, which could bypass the limits in the electrolyte conduction. This reduces the total cell impedance and, thus, the lower cell voltage is reached later. The different diffusion paths are visualized in \cref{fig:resolved_fast_slow_diffusion} where the flux by diffusion in the cathode, i.e., $\vec{N}_c = -D \nabla c$ is compared for different values of the diffusion coefficient within the resolved model.
\begin{figure}[ht]
  \centering
  \begin{subfigure}[b]{0.45\textwidth}
    \centering
    \includegraphics[width=0.75\textwidth]{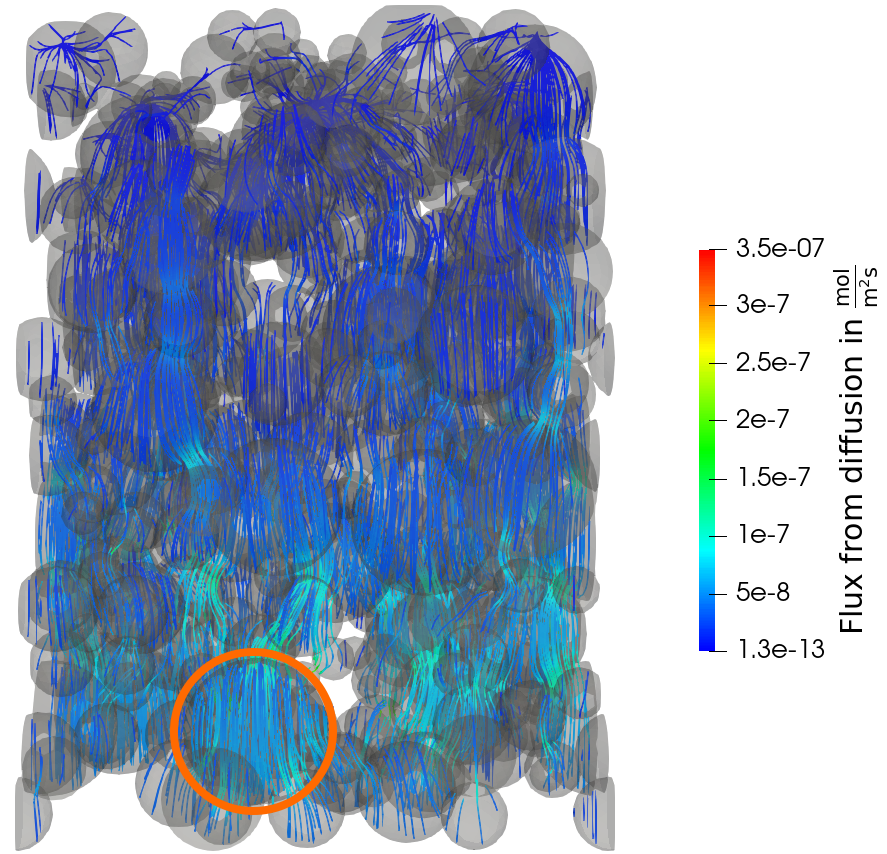}
    \caption{Fast diffusion ($D_\text{fast} = 10^{-12} \frac{\text{m}^2}{\text{s}}$) with a dominating inter-particle flux.}
    \label{fig:resolved_fast_diffusion}
  \end{subfigure}
  \hfill
  \begin{subfigure}[b]{0.45\textwidth}
    \centering
    \includegraphics[width=0.75\textwidth]{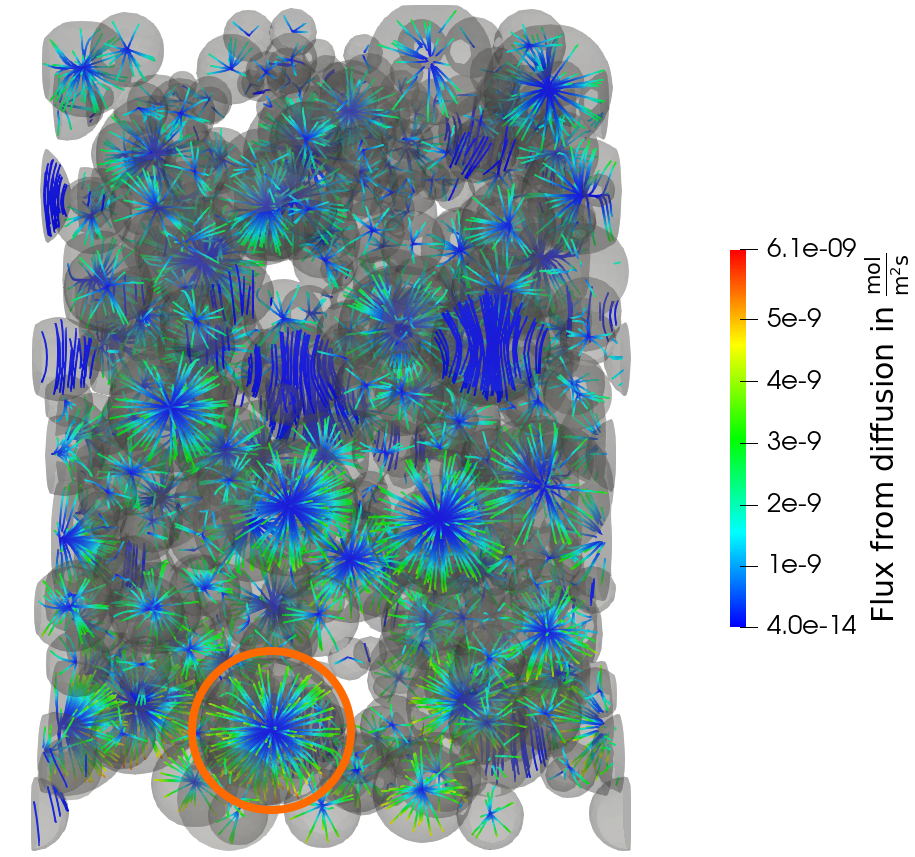}
    \caption{Slow diffusion ($D_\text{slow} = 10^{-15} \frac{\text{m}^2}{\text{s}}$) with a dominating radial flux within the particles.}
    \label{fig:resolved_slow_diffusion}
  \end{subfigure}
  \caption{Comparison of the effect of fast and slow diffusion within the resolved model at the end of discharge. The orange circles denote the fundamentally different diffusion paths within the same particle. Note the different ranges of the color bars.}
  \label{fig:resolved_fast_slow_diffusion}
\end{figure}
In \cref{fig:resolved_fast_diffusion}, the value of the diffusion coefficient is set to $D_\text{fast} = 10^{-12} \frac{\text{m}^2}{\text{s}}$, and in \cref{fig:resolved_slow_diffusion}, it is set to $D_\text{slow} = 10^{-15} \frac{\text{m}^2}{\text{s}}$. In the case of fast diffusion, an inter-particle flux is visible, while in the case of slow diffusion, the flux is mainly radial within each particle. The inter-particle flux in the case of large diffusion coefficients cannot be resolved by the P2D model, and thus, deviations between both models occur for large values of the diffusion coefficient.\\
Second, a different slope of the SOC for small diffusion coefficients becomes visible. In this region of the diffusion coefficient, diffusion is the limiting transport process. Within the resolved model, the orientation, intersection, and different sizes of the active material particles are considered, leading to complex diffusion paths. Within the P2D model, the diffusion is always spatially homogeneous and perfectly radial within every particle, and the entire surface of a particle serves as an intercalation surface, leading to a homogeneous concentration on the surface and, therefore, smaller gradients in the concentration. These aspects reduce the transport resistance originating from diffusion, and thus, the overall cell impedance is underestimated compared to the resolved model. Therefore, the lower cell voltage is reached later, and the SOC at the end of discharge is smaller. A visual quantification of the deviation between both models and an identification of regions where the P2D model over- or underestimates the SOC is given in \cref{fig:comparison_resolved_homogenized_diffusion_result_diagonal}. There, one cross represents one value of the diffusion coefficient. Again, the SOC at the end of discharge computed with the resolved model is aligned on the horizontal axis and with the P2D model on the vertical axis. As seen before, the P2D model overestimates the SOC at the end of discharge for large values of the diffusion coefficient and underestimates it for small diffusion coefficients. For larger values of the diffusion coefficient, the marks are almost at the same position, indicating that a change in the diffusion coefficient does not change the result anymore due to other limiting transport mechanisms. From this study, it can be concluded and quantified that the P2D model misses two significant effects: The inter-particle diffusion for large diffusion coefficients and the additional resistances originating from diffusion in geometrically complex microstructures for small diffusion coefficients.\\ 
Now, the ionic conductivity in the solid electrolyte is varied within the bounds given in \cref{table:comparison_resolved_homogenized_intervals}, while the diffusion coefficient $D = 3.16 \cdot 10^{-14} \frac{\text{m}^2}{\text{s}}$, the electronic conductivity $\sigma = 0.32 \frac{\text{S}}{\text{m}}$, and the current $N_\rho = 6.27 \frac{\text{A}}{\text{m}^2}$, are fixed to their averaged values. In \cref{fig:comparison_resolved_homogenized_el_cond_result}, the SOC at the end of discharge computed with the resolved model and with the P2D model are shown.
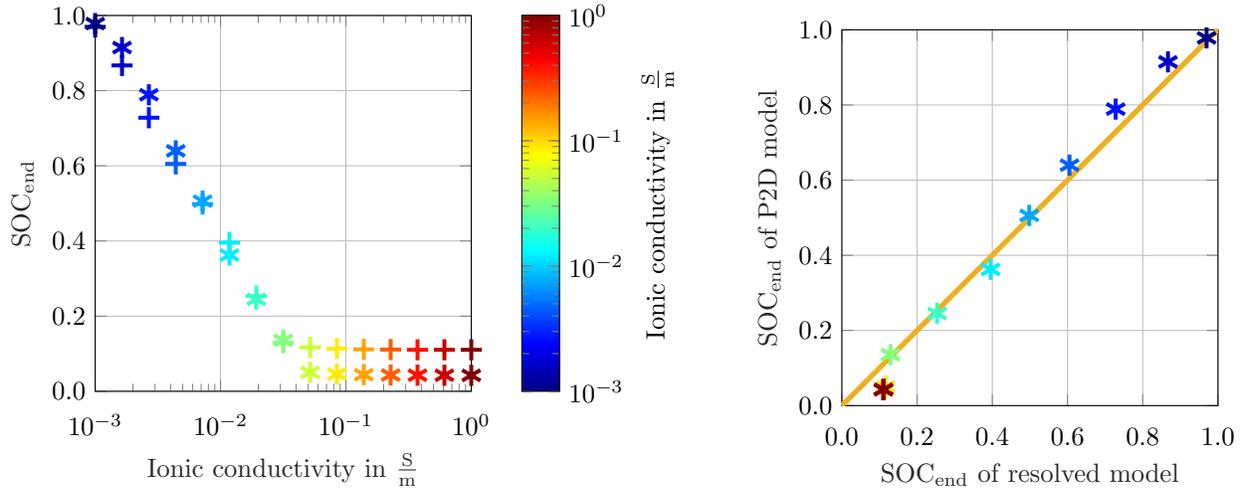
\begin{figure}[ht]
  \centering
  \begin{subfigure}[b]{0.4\textwidth}
    \centering
%
%
\definecolor{mycolor1}{rgb}{0.00000,0.00000,0.51562}%
\definecolor{mycolor2}{rgb}{0.00000,0.00000,0.79688}%
\definecolor{mycolor3}{rgb}{0.00000,0.07812,1.00000}%
\definecolor{mycolor4}{rgb}{0.00000,0.37500,1.00000}%
\definecolor{mycolor5}{rgb}{0.00000,0.65625,1.00000}%
\definecolor{mycolor6}{rgb}{0.00000,0.93750,1.00000}%
\definecolor{mycolor7}{rgb}{0.21875,1.00000,0.78125}%
\definecolor{mycolor8}{rgb}{0.51562,1.00000,0.48438}%
\definecolor{mycolor9}{rgb}{0.79688,1.00000,0.20312}%
\definecolor{mycolor10}{rgb}{1.00000,0.92188,0.00000}%
\definecolor{mycolor11}{rgb}{1.00000,0.64062,0.00000}%
\definecolor{mycolor12}{rgb}{1.00000,0.35938,0.00000}%
\definecolor{mycolor13}{rgb}{1.00000,0.06250,0.00000}%
\definecolor{mycolor14}{rgb}{0.78125,0.00000,0.00000}%
\definecolor{mycolor15}{rgb}{0.50000,0.00000,0.00000}%
\begin{tikzpicture}

\begin{axis}[%
width=5.0cm,
height=5.0cm,
point meta min=1e-3,
point meta max=1e0,
scale only axis,
xmode=log,
xmin=1e-3,
xmax=1e0,
x tick label style={
/pgf/number format/.cd,
precision=0,
/tikz/.cd,
yshift=-.5em},
xlabel style={font=\color{white!15!black}},
xlabel={Ionic conductivity in $\frac{\text{S}}{\text{m}}$},
ymin=0,
ymax=1,
y tick label style={
/pgf/number format/.cd,
fixed,
fixed zerofill,
precision=1,
/tikz/.cd},
ylabel style={font=\color{white!15!black}},
ylabel={$\text{SOC}_\text{end}$},
axis background/.style={fill=white},
xmajorgrids,
ymajorgrids,
colorbar,
colorbar style={
    ylabel=Ionic conductivity in $\frac{\text{S}}{\text{m}}$,
    ymode=log,
    yticklabel style={
        /pgf/number format/.cd,
        fixed,
        precision=0,
    }
},
colormap/jet
]
\addplot [color=mycolor1, line width=1.5pt, only marks, mark size=4.0pt, mark=asterisk, mark options={solid, mycolor1}, forget plot]
  table[row sep=crcr]{%
0.001	0.978525344054435\\
};
\addplot [color=mycolor1, line width=1.5pt, only marks, mark size=4.0pt, mark=+, mark options={solid, mycolor1}, forget plot]
  table[row sep=crcr]{%
0.001	0.969681514769729\\
};
\addplot [color=mycolor2, line width=1.5pt, only marks, mark size=4.0pt, mark=asterisk, mark options={solid, mycolor2}, forget plot]
  table[row sep=crcr]{%
0.00163789370695406	0.914672824489227\\
};
\addplot [color=mycolor2, line width=1.5pt, only marks, mark size=4.0pt, mark=+, mark options={solid, mycolor2}, forget plot]
  table[row sep=crcr]{%
0.00163789370695406	0.867076707667376\\
};
\addplot [color=mycolor3, line width=1.5pt, only marks, mark size=4.0pt, mark=asterisk, mark options={solid, mycolor3}, forget plot]
  table[row sep=crcr]{%
0.00268269579527972	0.789017515860816\\
};
\addplot [color=mycolor3, line width=1.5pt, only marks, mark size=4.0pt, mark=+, mark options={solid, mycolor3}, forget plot]
  table[row sep=crcr]{%
0.00268269579527972	0.727962089181345\\
};
\addplot [color=mycolor4, line width=1.5pt, only marks, mark size=4.0pt, mark=asterisk, mark options={solid, mycolor4}, forget plot]
  table[row sep=crcr]{%
0.00439397056076079	0.639937215702948\\
};
\addplot [color=mycolor4, line width=1.5pt, only marks, mark size=4.0pt, mark=+, mark options={solid, mycolor4}, forget plot]
  table[row sep=crcr]{%
0.00439397056076079	0.605213577740531\\
};
\addplot [color=mycolor5, line width=1.5pt, only marks, mark size=4.0pt, mark=asterisk, mark options={solid, mycolor5}, forget plot]
  table[row sep=crcr]{%
0.00719685673001152	0.50612614017376\\
};
\addplot [color=mycolor5, line width=1.5pt, only marks, mark size=4.0pt, mark=+, mark options={solid, mycolor5}, forget plot]
  table[row sep=crcr]{%
0.00719685673001152	0.498202141613375\\
};
\addplot [color=mycolor6, line width=1.5pt, only marks, mark size=4.0pt, mark=asterisk, mark options={solid, mycolor6}, forget plot]
  table[row sep=crcr]{%
0.0117876863479359	0.362465226702971\\
};
\addplot [color=mycolor6, line width=1.5pt, only marks, mark size=4.0pt, mark=+, mark options={solid, mycolor6}, forget plot]
  table[row sep=crcr]{%
0.0117876863479359	0.395597036824926\\
};
\addplot [color=mycolor7, line width=1.5pt, only marks, mark size=4.0pt, mark=asterisk, mark options={solid, mycolor7}, forget plot]
  table[row sep=crcr]{%
0.0193069772888325	0.245144503194716\\
};
\addplot [color=mycolor7, line width=1.5pt, only marks, mark size=4.0pt, mark=+, mark options={solid, mycolor7}, forget plot]
  table[row sep=crcr]{%
0.0193069772888325	0.253363954207447\\
};
\addplot [color=mycolor8, line width=1.5pt, only marks, mark size=4.0pt, mark=asterisk, mark options={solid, mycolor8}, forget plot]
  table[row sep=crcr]{%
0.0316227766016838	0.135592450008328\\
};
\addplot [color=mycolor8, line width=1.5pt, only marks, mark size=4.0pt, mark=+, mark options={solid, mycolor8}, forget plot]
  table[row sep=crcr]{%
0.0316227766016838	0.129354213573001\\
};
\addplot [color=mycolor9, line width=1.5pt, only marks, mark size=4.0pt, mark=asterisk, mark options={solid, mycolor9}, forget plot]
  table[row sep=crcr]{%
0.0517947467923121	0.0503652476431774\\
};
\addplot [color=mycolor9, line width=1.5pt, only marks, mark size=4.0pt, mark=+, mark options={solid, mycolor9}, forget plot]
  table[row sep=crcr]{%
0.0517947467923121	0.116764352728037\\
};
\addplot [color=mycolor10, line width=1.5pt, only marks, mark size=4.0pt, mark=asterisk, mark options={solid, mycolor10}, forget plot]
  table[row sep=crcr]{%
0.0848342898244072	0.0457470085941487\\
};
\addplot [color=mycolor10, line width=1.5pt, only marks, mark size=4.0pt, mark=+, mark options={solid, mycolor10}, forget plot]
  table[row sep=crcr]{%
0.0848342898244072	0.113595300409895\\
};
\addplot [color=mycolor11, line width=1.5pt, only marks, mark size=4.0pt, mark=asterisk, mark options={solid, mycolor11}, forget plot]
  table[row sep=crcr]{%
0.138949549437314	0.04414259559328\\
};
\addplot [color=mycolor11, line width=1.5pt, only marks, mark size=4.0pt, mark=+, mark options={solid, mycolor11}, forget plot]
  table[row sep=crcr]{%
0.138949549437314	0.111696569486485\\
};
\addplot [color=mycolor12, line width=1.5pt, only marks, mark size=4.0pt, mark=asterisk, mark options={solid, mycolor12}, forget plot]
  table[row sep=crcr]{%
0.227584592607479	0.0433387347085242\\
};
\addplot [color=mycolor12, line width=1.5pt, only marks, mark size=4.0pt, mark=+, mark options={solid, mycolor12}, forget plot]
  table[row sep=crcr]{%
0.227584592607479	0.111062288448945\\
};
\addplot [color=mycolor13, line width=1.5pt, only marks, mark size=4.0pt, mark=asterisk, mark options={solid, mycolor13}, forget plot]
  table[row sep=crcr]{%
0.372759372031494	0.0428969122596444\\
};
\addplot [color=mycolor13, line width=1.5pt, only marks, mark size=4.0pt, mark=+, mark options={solid, mycolor13}, forget plot]
  table[row sep=crcr]{%
0.372759372031494	0.110431023961174\\
};
\addplot [color=mycolor14, line width=1.5pt, only marks, mark size=4.0pt, mark=asterisk, mark options={solid, mycolor14}, forget plot]
  table[row sep=crcr]{%
0.610540229658533	0.0426466384503114\\
};
\addplot [color=mycolor14, line width=1.5pt, only marks, mark size=4.0pt, mark=+, mark options={solid, mycolor14}, forget plot]
  table[row sep=crcr]{%
0.610540229658533	0.110431400798825\\
};
\addplot [color=mycolor15, line width=1.5pt, only marks, mark size=4.0pt, mark=asterisk, mark options={solid, mycolor15}, forget plot]
  table[row sep=crcr]{%
1	0.042456161062122\\
};
\addplot [color=mycolor15, line width=1.5pt, only marks, mark size=4.0pt, mark=+, mark options={solid, mycolor15}, forget plot]
  table[row sep=crcr]{%
1	0.110430897956964\\
};
\end{axis}

\end{tikzpicture}%
    \caption{SOC at the end of discharge with the resolved (+) and the P2D model ($\ast$).}
    \label{fig:comparison_resolved_homogenized_el_cond_result}
  \end{subfigure}
  \hfill
  \begin{subfigure}[b]{0.4\textwidth}
    \centering
%
%
\definecolor{mycolor1}{rgb}{0.00000,0.00000,0.51562}%
\definecolor{mycolor2}{rgb}{0.00000,0.00000,0.79688}%
\definecolor{mycolor3}{rgb}{0.00000,0.07812,1.00000}%
\definecolor{mycolor4}{rgb}{0.00000,0.37500,1.00000}%
\definecolor{mycolor5}{rgb}{0.00000,0.65625,1.00000}%
\definecolor{mycolor6}{rgb}{0.00000,0.93750,1.00000}%
\definecolor{mycolor7}{rgb}{0.21875,1.00000,0.78125}%
\definecolor{mycolor8}{rgb}{0.51562,1.00000,0.48438}%
\definecolor{mycolor9}{rgb}{0.79688,1.00000,0.20312}%
\definecolor{mycolor10}{rgb}{1.00000,0.92188,0.00000}%
\definecolor{mycolor11}{rgb}{1.00000,0.64062,0.00000}%
\definecolor{mycolor12}{rgb}{1.00000,0.35938,0.00000}%
\definecolor{mycolor13}{rgb}{1.00000,0.06250,0.00000}%
\definecolor{mycolor14}{rgb}{0.78125,0.00000,0.00000}%
\definecolor{mycolor15}{rgb}{0.50000,0.00000,0.00000}%
\definecolor{mycolor16}{rgb}{0.92900,0.69400,0.12500}%
\begin{tikzpicture}

\begin{axis}[%
width=5.0cm,
height=5.0cm,
point meta min=-3.0,
point meta max=0.0,
scale only axis,
xmin=0,
xmax=1,
x tick label style={
/pgf/number format/.cd,
fixed,
fixed zerofill,
precision=1,
/tikz/.cd,
yshift=-.5em},
xlabel style={font=\color{white!15!black}},
xlabel={$\text{SOC}_\text{end}$ of resolved model},
ymin=0,
ymax=1,
y tick label style={
/pgf/number format/.cd,
fixed,
fixed zerofill,
precision=1,
/tikz/.cd},
ylabel style={font=\color{white!15!black}},
ylabel={$\text{SOC}_\text{end}$ of P2D model},
axis background/.style={fill=white},
xmajorgrids,
ymajorgrids
]
\addplot [color=mycolor1, line width=1.5pt, only marks, mark size=4.0pt, mark=asterisk, mark options={solid, mycolor1}, forget plot]
  table[row sep=crcr]{%
0.969681514769729	0.978525344054435\\
};
\addplot [color=mycolor2, line width=1.5pt, only marks, mark size=4.0pt, mark=asterisk, mark options={solid, mycolor2}, forget plot]
  table[row sep=crcr]{%
0.867076707667376	0.914672824489227\\
};
\addplot [color=mycolor3, line width=1.5pt, only marks, mark size=4.0pt, mark=asterisk, mark options={solid, mycolor3}, forget plot]
  table[row sep=crcr]{%
0.727962089181345	0.789017515860816\\
};
\addplot [color=mycolor4, line width=1.5pt, only marks, mark size=4.0pt, mark=asterisk, mark options={solid, mycolor4}, forget plot]
  table[row sep=crcr]{%
0.605213577740531	0.639937215702948\\
};
\addplot [color=mycolor5, line width=1.5pt, only marks, mark size=4.0pt, mark=asterisk, mark options={solid, mycolor5}, forget plot]
  table[row sep=crcr]{%
0.498202141613375	0.50612614017376\\
};
\addplot [color=mycolor6, line width=1.5pt, only marks, mark size=4.0pt, mark=asterisk, mark options={solid, mycolor6}, forget plot]
  table[row sep=crcr]{%
0.395597036824926	0.362465226702971\\
};
\addplot [color=mycolor7, line width=1.5pt, only marks, mark size=4.0pt, mark=asterisk, mark options={solid, mycolor7}, forget plot]
  table[row sep=crcr]{%
0.253363954207447	0.245144503194716\\
};
\addplot [color=mycolor8, line width=1.5pt, only marks, mark size=4.0pt, mark=asterisk, mark options={solid, mycolor8}, forget plot]
  table[row sep=crcr]{%
0.129354213573001	0.135592450008328\\
};
\addplot [color=mycolor9, line width=1.5pt, only marks, mark size=4.0pt, mark=asterisk, mark options={solid, mycolor9}, forget plot]
  table[row sep=crcr]{%
0.116764352728037	0.0503652476431774\\
};
\addplot [color=mycolor10, line width=1.5pt, only marks, mark size=4.0pt, mark=asterisk, mark options={solid, mycolor10}, forget plot]
  table[row sep=crcr]{%
0.113595300409895	0.0457470085941487\\
};
\addplot [color=mycolor11, line width=1.5pt, only marks, mark size=4.0pt, mark=asterisk, mark options={solid, mycolor11}, forget plot]
  table[row sep=crcr]{%
0.111696569486485	0.04414259559328\\
};
\addplot [color=mycolor12, line width=1.5pt, only marks, mark size=4.0pt, mark=asterisk, mark options={solid, mycolor12}, forget plot]
  table[row sep=crcr]{%
0.111062288448945	0.0433387347085242\\
};
\addplot [color=mycolor13, line width=1.5pt, only marks, mark size=4.0pt, mark=asterisk, mark options={solid, mycolor13}, forget plot]
  table[row sep=crcr]{%
0.110431023961174	0.0428969122596444\\
};
\addplot [color=mycolor14, line width=1.5pt, only marks, mark size=4.0pt, mark=asterisk, mark options={solid, mycolor14}, forget plot]
  table[row sep=crcr]{%
0.110431400798825	0.0426466384503114\\
};
\addplot [color=mycolor15, line width=1.5pt, only marks, mark size=4.0pt, mark=asterisk, mark options={solid, mycolor15}, forget plot]
  table[row sep=crcr]{%
0.110430897956964	0.042456161062122\\
};
\addplot [color=mycolor16, line width=2.0pt, forget plot]
  table[row sep=crcr]{%
0	0\\
1	1\\
};
\end{axis}

\end{tikzpicture}%
    \caption{Comparison of the SOC at the end of discharge computed with both models.}
    \label{fig:comparison_resolved_homogenized_el_cond_diagonal}
  \end{subfigure}
  \caption{Comparison of the SOC at the end of discharge for different values of the ionic conductivity in the solid electrolyte.}
\end{figure}
Again, the progression of both models follows a similar trend. The SOC at the end of discharge approaches a constant value for large values of the ionic conductivity indicating other transport phenomena to be limiting in this region. For small values of the ionic conductivity, the SOC approaches one, meaning that no charge is transported. However, an offset is present for large values of the ionic conductivity between the SOC computed with both models. In this case, this is attributed to an insufficient homogenization strategy, as the effective transport properties of the P2D model do not match the actual transport properties in the resolved model. For small values of the ionic conductivity, the small ionic conduction cannot be compensated by diffusion between the particles as inter-particle diffusion is not possible in the P2D model, resulting in less transported charge. For further comparison, the SOC at the end of discharge computed with the resolved model is again compared in a diagram where the results of the resolved model are aligned on the horizontal axis and of the P2D model on the vertical axis (c.f.\cref{fig:comparison_resolved_homogenized_el_cond_diagonal}). Regions can be identified where the P2D model overestimates the SOC (i.e., at small ionic conductivities) and regions where it underestimates the SOC (i.e., at large ionic conductivities), as discussed before. \\
For another study, the electronic conductivity~$\sigma = 0.32 \frac{\text{S}}{\text{m}}$, and the current~$N_\rho = 6.27 \frac{\text{A}}{\text{m}^2}$ are fixed, while the ionic conductivity and the diffusion coefficient are varied within the bounds given in \cref{table:comparison_resolved_homogenized_intervals}. The results computed with the resolved model are shown in \cref{fig:comparison_resolved_homogenized_two_parameters_plane_resolved} by representing the computed SOC at the end of discharge as a plane depending on both parameters.
\begin{figure}[ht]
  \centering
  \begin{subfigure}[b]{0.45\textwidth}
    \centering
    \includegraphics{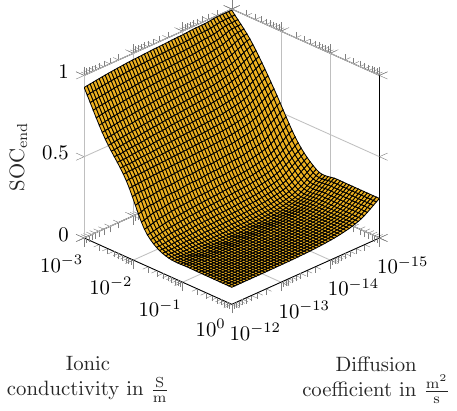}
    \caption{SOC at the end of discharge computed with the resolved model.}
    \label{fig:comparison_resolved_homogenized_two_parameters_plane_resolved}
  \end{subfigure}
  \hfill
  \begin{subfigure}[b]{0.45\textwidth}
    \centering
    \includegraphics{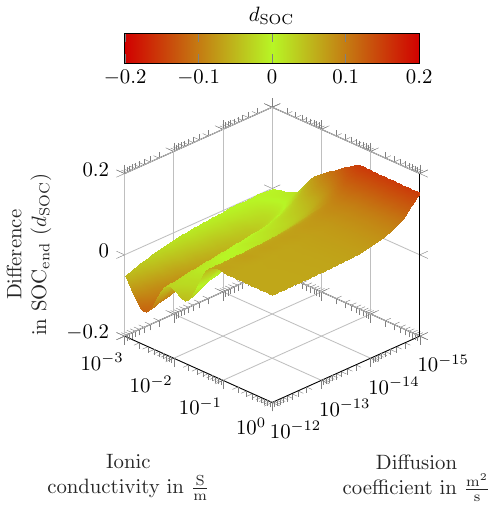}
    \caption{Difference in the SOC at the end of discharge between both models.}
    \label{fig:comparison_resolved_homogenized_two_parameters_difference}
  \end{subfigure}
  \caption{Results for~$\text{SOC}_\text{end}$ for different values of the ionic conductivity and the diffusion coefficient.}
\end{figure}
The difference in the SOC between the models $d_\text{SOC} = \text{SOC}_\text{end,res} - \text{SOC}_\text{end,P2D}$ is shown in \cref{fig:comparison_resolved_homogenized_two_parameters_difference}. Differences of $|d_\text{SOC}| \approx 0.15$ become visible. These deviations are local over- as well as underestimations. While a constant difference could still be corrected by an update of the homogenization parameters, a correction of the homogenization parameters is non-trivial in this case. The local sensitivity of this deviation can be a measure to identify regions in which the results of the P2D model are insensitive to the input parameters and regions in which the results are sensitive and only meaningful within a small interval close to the calibration point. The local sensitivity is quantified by the magnitude of the local derivative of~$d_\text{SOC}$, $m_\text{SOC} = \sqrt{\left(\parder{d_\text{SOC}}{\kappa} \kappa_1 \right)^2 + \left(\parder{d_\text{SOC}}{D} D_1\right)^2}$. It is shown in \cref{fig:comparison_resolved_homogenized_two_parameters_difference_derivative}.
\begin{figure}[ht]
  \centering
  \includegraphics{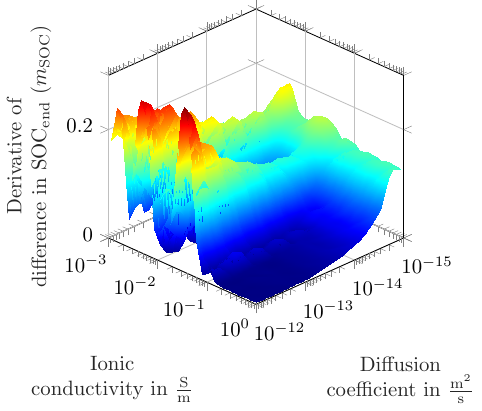}
  \caption{Derivative of the difference in the SOC at the end of discharge for different values of the diffusion coefficient and the ionic conductivity.}
  \label{fig:comparison_resolved_homogenized_two_parameters_difference_derivative}
\end{figure}
Regions are visible in which the value of the derivative is almost zero. In this region (large ionic conductivity and small diffusion coefficient), the P2D model may be easily used. In the other regions, the sensitivity is non-zero, and the P2D model leads to significant deviations that cannot be compensated by a constant offset. There, a P2D model gives reliable results only close to the calibration point and has only a limited predictive character.
\subsubsection{Comparison of computational costs}
The computational costs of a simulation run may be expressed by the total CPU time, which is defined as $t_\text{CPU} = n_\text{CPU} \, t_\text{run}$, with the number of used CPUs~$n_\text{CPU}$ and the total run time~$t_\text{run}$. All computations were performed on similar, i.e., only slightly different hardware using the software project \textit{4C multiphysics}~\cite{4C}. Within the resolved model, the linear system of equations contains $\approx$1.8 million unknowns, while the P2D model contains $\approx\,$600 unknowns. For both models, this linear system of equations needs to be solved multiple times within each time step to solve the nonlinear system of equations by the Newton-Raphson iteration. The mean values for the solution of the resolved model and of the P2D model for the evaluations of 150 samples within the range in \cref{table:comparison_resolved_homogenized_intervals} are listed in \cref{table:comparison_run_time}.
\begin{table}[ht]
  \renewcommand{\arraystretch}{1.2}
  \centering
  \caption{Averaged values of the computational costs for the evaluation of 150 samples from the range in \cref{table:comparison_resolved_homogenized_intervals}.}
  \begin{tabular}{| c | c | c |}
    \hline
    \textbf{quantity}             & \textbf{value resolved} & \textbf{value P2D} \\
    \hline
    number of CPUs~($n_\text{CPU}$) & 48 & 1\\
    run time~($t_\text{run}$)       & 32.8 h & 90 s \\
    CPU time~($t_\text{CPU}$)       & 1601.7 h & 90 s \\
    \hline
  \end{tabular}
  \label{table:comparison_run_time}
\end{table}
Obviously, the solution of the P2D model is drastically faster than the solution of the resolved model. It should be noted that the computational implementation of both models is not optimized for performance. Therefore, the listed values of computational times could be smaller using optimized implementations, but the overall picture would not change.
\FloatBarrier
\section{Conclusion}
Within this study, the abilities of a geometrically three-dimensionally resolved model and a P2D model to predict the charge and discharge behavior of a solid-state battery cell in a global sense, i.e., for a wide range of various input parameters, are analyzed, quantified, and compared. The charge and discharge behavior deviates, especially for extreme values of the diffusion coefficient. The reason for this is that the P2D model cannot capture inter-particle diffusion as well as an inhomogeneous intra-particle diffusion. This leads to deviations between crucial output quantities of the models, e.g., the SOC when reaching a certain cell voltage with the same input parameters. The deviations are further quantified by comparing the global and local sensitivity of the output quantities on the input parameters. As a measure of the global sensitivity, the Sobol indices of first, second, and total order are used. The indices reveal a non-negligible deviation between the sensitivities computed with both models, which is confirmed by analyzing the local sensitivity. Especially the different global and local sensitivities of the models on the input parameters question the ability to predict the discharge behavior of the SSB cell with a P2D model in a wide range of input parameters. Certainly, multiple extensions or modifications of the homogenization procedure within the P2D model exist to improve selected results. However, an a priori knowledge on when to enable which extension is, in general, not given. Besides these deviations, local quantities such as local flux or stress maxima are not even resolved within the P2D model. These quantities are often the decisive factor when evaluating a cell concept, as they could be the origin of degradation or the bottleneck for fast transportation.\\
We conclude that using P2D models should be given special attention, especially when exploring ranges of parameter values for which no data for comparison is available. Certainly, the P2D model provides a tremendous reduction of the computational costs compared to the fully resolved model. Considering both aspects, we recommend using the P2D model to quickly obtain results in a wide range of input parameters and to compare its results at selected points with the solution of a fully resolved model and correct the results, or even adapt the formulation of the P2D model if the deviation is unacceptable. A theoretically profound approach to combine the accuracy of a fully resolved model and the computational cheapness of a P2D model is given by learning their relationship using multi-fidelity methods as done, for example, in~\cite{Nitzler2022} for fluid simulations.
\FloatBarrier
\section*{Funding}
We gratefully acknowledge support by the Bavarian Ministry of Economic Affairs, Regional Development and Energy [project ``Industrialisierbarkeit von Festk\"orperelektrolytzellen''] and the German Federal Ministry of Education and Research [FestBatt~2 (03XP0435B)].
\FloatBarrier
\begin{appendices}
    \section{Material parameters}
\label{sec:material_parameters}
\begin{figure}[ht]
    \centering
    \begin{subfigure}[b]{0.45\textwidth}
        \centering
        \input{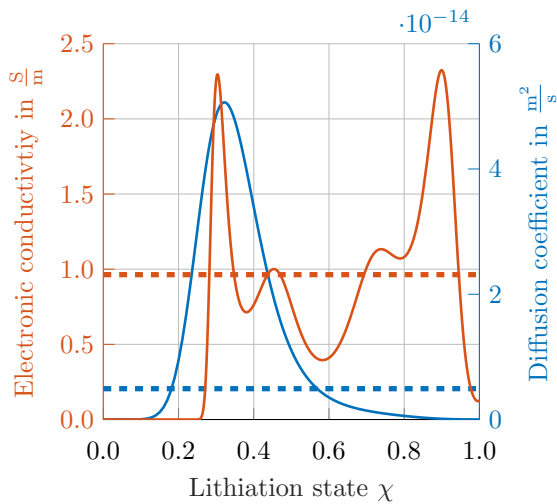}
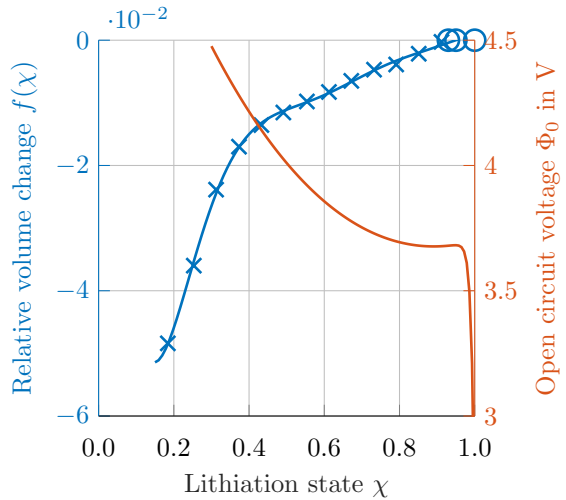
        \caption{Lithiation dependent transport properties of NMC-622. The dashed lines represent the average value between~$\chi_\text{min}$ and~$\chi_\text{max}$, i.e., $\bar{\sigma} = 0.205 \frac{\text{S}}{\text{m}}$ and $\bar{D} = 2.313 \cdot 10^{-14}  \frac{\text{m}^2}{\text{s}}$~\cite{Neumann2020}.}
        \label{fig:NMC_transport_properties}
    \end{subfigure}
    \hfill
    \begin{subfigure}[b]{0.45\textwidth}
        \centering
%
%
\definecolor{mycolor1}{rgb}{0.00000,0.44700,0.74100}%
\definecolor{mycolor2}{rgb}{0.85000,0.32500,0.09800}%
\begin{tikzpicture}

\begin{axis}[%
width=5.0cm,
height=5.0cm,
scale only axis,
xmin=0,
xmax=1,
x tick label style={
/pgf/number format/.cd,
fixed,
fixed zerofill,
precision=1,
/tikz/.cd,
yshift=-.5em},
xlabel style={font=\color{white!15!black}},
xlabel={Lithiation state $\chi$},
every outer y axis line/.append style={mycolor1},
every y tick label/.append style={font=\color{mycolor1}},
every y tick/.append style={mycolor1},
ymin=-0.06,
ymax=0.0,
y tick label style={
/pgf/number format/.cd,
fixed,
fixed zerofill,
precision=0,
/tikz/.cd},
ylabel style={font=\color{mycolor1}},
ylabel={Relative volume change $f(\chi)$},
axis background/.style={fill=white},
axis x line*=bottom,
axis y line*=left,
xmajorgrids,
ymajorgrids
]
\addplot [color=mycolor1, line width=1.0pt, only marks, mark size=4.0pt, mark=x, mark options={solid, mycolor1}, forget plot]
  table[row sep=crcr]{%
0.184036	-0.0484\\
0.253076	-0.035991\\
0.312314	-0.023856\\
0.373186	-0.016995\\
0.432802	-0.013548\\
0.491154	-0.011543\\
0.553774	-0.009782\\
0.612359	-0.008266\\
0.671865	-0.006491\\
0.732295	-0.004702\\
0.791112	-0.003878\\
0.851196	-0.00216\\
0.911624	-0.00027\\
};
\addplot [color=mycolor1, line width=1.0pt, only marks, mark size=4.0pt, mark=o, mark options={solid, mycolor1}, forget plot]
  table[row sep=crcr]{%
0.93	0\\
0.95	0\\
1	0\\
1.05	0\\
};
\addplot [color=mycolor1, line width=1.0pt, forget plot]
  table[row sep=crcr]{%
0.15	-0.0514334788613085\\
0.15959595959596	-0.0510899509352353\\
0.169191919191919	-0.0503299826140962\\
0.178787878787879	-0.0492290304465524\\
0.188383838383838	-0.0478545340034029\\
0.197979797979798	-0.0462664747501269\\
0.207575757575758	-0.0445179108762336\\
0.217171717171717	-0.042655488573224\\
0.226767676767677	-0.0407199302529682\\
0.236363636363636	-0.0387465001983058\\
0.245959595959596	-0.0367654481376699\\
0.255555555555556	-0.0348024312355449\\
0.265151515151515	-0.0328789149905589\\
0.274747474747475	-0.0310125535330156\\
0.284343434343434	-0.0292175498136751\\
0.293939393939394	-0.0275049961755823\\
0.303535353535354	-0.0258831958007509\\
0.313131313131313	-0.0243579655235096\\
0.322727272727273	-0.0229329205023106\\
0.332323232323232	-0.021609741241811\\
0.341919191919192	-0.0203884234570254\\
0.351515151515152	-0.0192675112713612\\
0.361111111111111	-0.0182443142403367\\
0.370707070707071	-0.017315108692789\\
0.38030303030303	-0.0164753238813754\\
0.38989898989899	-0.015719713434176\\
0.39949494949495	-0.0150425125991941\\
0.409090909090909	-0.0144375817735766\\
0.418686868686869	-0.0138985368093344\\
0.428282828282828	-0.0134188665873915\\
0.437878787878788	-0.0129920383517517\\
0.447474747474747	-0.0126115912955978\\
0.457070707070707	-0.012271218891118\\
0.466666666666667	-0.0119648404548748\\
0.476262626262626	-0.0116866624405163\\
0.485858585858586	-0.0114312299506349\\
0.495454545454545	-0.0111934689595711\\
0.505050505050505	-0.010968719738987\\
0.514646464646465	-0.010752761977995\\
0.524242424242424	-0.0105418320896505\\
0.533838383838384	-0.0103326331956233\\
0.543434343434343	-0.010122338280838\\
0.553030303030303	-0.00990858700990695\\
0.562626262626263	-0.00968947669713428\\
0.572222222222222	-0.00946354792192917\\
0.581818181818182	-0.00922976528139402\\
0.591414141414141	-0.00898749377193643\\
0.601010101010101	-0.00873647129166367\\
0.610606060606061	-0.0084767777554107\\
0.62020202020202	-0.00820880131416493\\
0.62979797979798	-0.00793320217072616\\
0.639393939393939	-0.00765087448336976\\
0.648989898989899	-0.00736290684937019\\
0.658585858585859	-0.00707054186014017\\
0.668181818181818	-0.00677513521979979\\
0.677777777777778	-0.00647811491902264\\
0.687373737373737	-0.00618094095589598\\
0.696969696969697	-0.00588506609566149\\
0.706565656565657	-0.00559189816110277\\
0.716161616161616	-0.00530276434539269\\
0.725757575757576	-0.00501887803921758\\
0.735353535353535	-0.00474130866398273\\
0.744949494949495	-0.00447095500287127\\
0.754545454545455	-0.00420852252162204\\
0.764141414141414	-0.00395450517074759\\
0.773737373737374	-0.00370917216110456\\
0.783333333333333	-0.0034725602045079\\
0.792929292929293	-0.00324447171127715\\
0.802525252525253	-0.00302447943647931\\
0.812121212121212	-0.00281193806671559\\
0.821717171717172	-0.00260600323915899\\
0.831313131313131	-0.00240565848476529\\
0.840909090909091	-0.00220975058743056\\
0.850505050505051	-0.00201703385080104\\
0.86010101010101	-0.0018262237647043\\
0.86969696969697	-0.00163606056287324\\
0.879292929292929	-0.00144538316381895\\
0.888888888888889	-0.00125321398668503\\
0.898484848484849	-0.00105885513383959\\
0.908080808080808	-0.000861996431987408\\
0.917676767676768	-0.000662835823719926\\
0.927272727272727	-0.000462212601220355\\
0.936868686868687	-0.000261753973898632\\
0.946464646464647	-6.40354618334088e-05\\
0.956060606060606	0.000127244393179765\\
0.965656565656566	0.000307074507253036\\
0.975252525252525	0.000468926432700759\\
0.984848484848485	0.000604533743524788\\
0.994444444444445	0.000703655135967846\\
1.0040404040404	0.00075382075241634\\
1.01363636363636	0.000740061237041192\\
1.02323232323232	0.000644619031143157\\
1.03282828282828	0.000446641416378471\\
1.04242424242424	0.000121854814327069\\
1.0520202020202	-0.000357780149712009\\
1.06161616161616	-0.00102443331024391\\
1.07121212121212	-0.00191478984448144\\
1.08080808080808	-0.00307045586616304\\
1.09040404040404	-0.00453838522232458\\
1.1	-0.00637132798470789\\
};

\end{axis}

\begin{axis}[%
width=5.0cm,
height=5.0cm,
scale only axis,
every outer y axis line/.append style={mycolor2},
every y tick label/.append style={font=\color{mycolor2}},
every y tick/.append style={mycolor2},
xmin=0,
xmax=1,
axis x line=none,
ymin=3.0,
ymax=4.5,
y tick label style={
/pgf/number format/.cd,
/tikz/.cd},
ylabel style={font=\color{mycolor2}},
ylabel={Open circuit voltage $\Phi_0$ in V},
axis x line*=bottom,
axis y line*=right
]
\addplot [color=mycolor2, line width=1.0pt, forget plot]
  table[row sep=crcr]{%
0.3 4.47619927240105\\
0.307070707070707 4.45582451376056\\
0.314141414141414 4.43574107056826\\
0.321212121212121 4.41594886951643\\
0.328282828282828 4.39644765830068\\
0.335353535353535 4.37723702746023\\
0.342424242424242 4.35831642956736\\
0.349494949494949 4.33968519611955\\
0.356565656565657 4.32134255243552\\
0.363636363636364 4.30328763081285\\
0.370707070707071 4.28551948216771\\
0.377777777777778 4.26803708634665\\
0.384848484848485 4.25083936127388\\
0.391919191919192 4.23392517107541\\
0.398989898989899 4.21729333330258\\
0.406060606060606 4.20094262536124\\
0.413131313131313 4.18487179023908\\
0.42020202020202 4.16907954161204\\
0.427272727272727 4.15356456840006\\
0.434343434343434 4.13832553883432\\
0.441414141414141 4.12336110408984\\
0.448484848484848 4.10866990153148\\
0.455555555555555 4.09425055761504\\
0.462626262626263 4.08010169048072\\
0.46969696969697 4.06622191227157\\
0.476767676767677 4.05260983120601\\
0.483838383838384 4.03926405342996\\
0.490909090909091 4.02618318467165\\
0.497979797979798 4.0133658317191\\
0.505050505050505 4.00081060373859\\
0.512121212121212 3.98851611345006\\
0.519191919191919 3.97648097817383\\
0.526262626262626 3.96470382076164\\
0.533333333333333 3.95318327042322\\
0.54040404040404 3.94191796345897\\
0.547474747474747 3.93090654390784\\
0.554545454545454 3.92014766411865\\
0.561616161616162 3.90963998525251\\
0.568686868686869 3.89938217772274\\
0.575757575757576 3.88937292157861\\
0.582828282828283 3.87961090683814\\
0.58989898989899 3.87009483377496\\
0.596969696969697 3.86082341316362\\
0.604040404040404 3.85179536648736\\
0.611111111111111 3.84300942611191\\
0.618181818181818 3.8344643354287\\
0.625252525252525 3.82615884897028\\
0.632323232323232 3.81809173250078\\
0.639393939393939 3.81026176308373\\
0.646464646464646 3.80266772912953\\
0.653535353535354 3.79530843042446\\
0.660606060606061 3.78818267814323\\
0.667676767676768 3.78128929484642\\
0.674747474747475 3.77462711446467\\
0.681818181818182 3.76819498227066\\
0.688888888888889 3.76199175484031\\
0.695959595959596 3.75601630000432\\
0.703030303030303 3.75026749679087\\
0.71010101010101 3.74474423536075\\
0.717171717171717 3.73944541693546\\
0.724242424242424 3.7343699537192\\
0.731313131313131 3.72951676881543\\
0.738383838383838 3.72488479613869\\
0.745454545454545 3.72047298032212\\
0.752525252525252 3.71628027662136\\
0.75959595959596 3.71230565081522\\
0.766666666666667 3.70854807910363\\
0.773737373737374 3.70500654800316\\
0.780808080808081 3.70168005424053\\
0.787878787878788 3.6985676046444\\
0.794949494949495 3.69566821603546\\
0.802020202020202 3.69298091511488\\
0.809090909090909 3.69050473835049\\
0.816161616161616 3.68823873185844\\
0.823232323232323 3.68618195127564\\
0.83030303030303 3.6843334616095\\
0.837373737373737 3.68269233703302\\
0.844444444444444 3.68125766054535\\
0.851515151515151 3.68002852330237\\
0.858585858585859 3.67900402313585\\
0.865656565656566 3.6781832610791\\
0.872727272727273 3.67756533299432\\
0.87979797979798 3.67714930916538\\
0.886868686868687 3.67693418432381\\
0.893939393939394 3.67691875503528\\
0.901010101010101 3.6771013186283\\
0.908080808080808 3.67747893369498\\
0.915151515151515 3.67804560348254\\
0.922222222222222 3.67878781309469\\
0.929292929292929 3.67967356566211\\
0.936363636363636 3.68062544709972\\
0.943434343434343 3.68145445308287\\
0.95050505050505 3.68169741857838\\
0.957575757575758 3.68021762288952\\
0.964646464646465 3.67422357599325\\
0.971717171717172 3.65685842139712\\
0.978787878787879 3.61127770022431\\
0.985858585858586 3.49609989452052\\
0.992929292929293 3.20966204045106\\
1 2.50220468352966\\
};
\end{axis}
\end{tikzpicture}%
        \caption{Open circuit voltage of NMC-622 (blue line)~\cite{Kremer2019} as a function of the lithiation state. Volume change of NMC-622 measured in~\cite{Koerver2018} as a function of the lithiation state. The crosses denote measured values, and the circles denote extrapolated values.}
        \label{fig:NMC_OCP_growth}
    \end{subfigure}
    \caption{Lithiation dependent material parameters of NMC.}
    \label{fig:NMC_lithiation_dependet_params}
\end{figure}

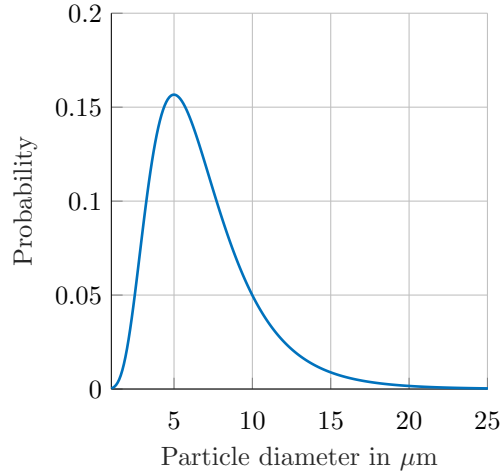
\begin{figure}[ht]
    \centering
%
%
\definecolor{mycolor1}{rgb}{0.00000,0.44700,0.74100}%
\definecolor{mycolor2}{rgb}{0.85000,0.32500,0.09800}%
\begin{tikzpicture}

\begin{axis}[%
width=5.0cm,
height=5.0cm,
scale only axis,
xmin=1,
xmax=25,
x tick label style={
/pgf/number format/.cd,
fixed,
fixed zerofill,
precision=0,
/tikz/.cd,
yshift=-.5em},
xlabel style={font=\color{white!15!black}},
xlabel={Particle diameter in $\mu$m},
ymin=0,
ymax=0.2,
y tick label style={
/pgf/number format/.cd,
fixed,
precision=2,
/tikz/.cd},
ylabel style={font=\color{white!15!black}},
ylabel={Probability},
axis background/.style={fill=white},
axis x line*=bottom,
axis y line*=left,
xmajorgrids,
ymajorgrids
]
\addplot [color=mycolor1, line width=1.0pt, forget plot]
  table[row sep=crcr]{%
1	0.000337106296024025\\
1.1	0.000683162373190857\\
1.2	0.00125375630221213\\
1.3	0.00212322394626806\\
1.4	0.00336539409915336\\
1.5	0.00504764072246458\\
1.6	0.00722583873278762\\
1.7	0.00994066504849316\\
1.8	0.0132154024034842\\
1.9	0.0170551862384658\\
2	0.0214474948761741\\
2.1	0.0263636119131817\\
2.2	0.0317607712811288\\
2.3	0.0375847125419728\\
2.4	0.0437724117134338\\
2.5	0.0502547997952006\\
2.6	0.0569593292505521\\
2.7	0.0638122930778436\\
2.8	0.0707408391771609\\
2.9	0.0776746534958914\\
3	0.0845473089849351\\
3.1	0.0912972943828465\\
3.2	0.0978687482310994\\
3.3	0.104211930361093\\
3.4	0.110283466395894\\
3.5	0.116046401483796\\
3.6	0.121470098296146\\
3.7	0.126530011904949\\
3.8	0.131207370998049\\
3.9	0.135488791363547\\
4	0.139365843952188\\
4.1	0.142834596296356\\
4.2	0.145895142752077\\
4.3	0.148551136011184\\
4.4	0.150809329643013\\
4.5	0.152679139080207\\
4.6	0.154172226454869\\
4.7	0.155302113000393\\
4.8	0.156083821334821\\
4.9	0.156533548803533\\
5	0.156668372151179\\
5.1	0.156505983084329\\
5.2	0.156064453748079\\
5.3	0.155362030744804\\
5.4	0.154416956047181\\
5.5	0.153247312978852\\
5.6	0.15187089533588\\
5.7	0.15030509768423\\
5.8	0.148566824879105\\
5.9	0.14667241889939\\
6	0.1446376011652\\
6.1	0.142477428600438\\
6.2	0.140206261809091\\
6.3	0.137837743848412\\
6.4	0.135384788200084\\
6.5	0.132859574658827\\
6.6	0.130273551974115\\
6.7	0.127637446193103\\
6.8	0.124961273760042\\
6.9	0.122254358528682\\
7	0.119525351938726\\
7.1	0.116782255695187\\
7.2	0.114032446370241\\
7.3	0.111282701421145\\
7.4	0.108539226184998\\
7.5	0.105807681471996\\
7.6	0.103093211433577\\
7.7	0.100400471430962\\
7.8	0.0977336556733069\\
7.9	0.095096524433575\\
8	0.0924924306845227\\
8.1	0.0899243460274246\\
8.2	0.0873948858125813\\
8.3	0.0849063333737016\\
8.4	0.0824606633182297\\
8.5	0.0800595638329024\\
8.6	0.0777044579785926\\
8.7	0.0753965239610607\\
8.8	0.0731367143748613\\
8.9	0.0709257744265643\\
9	0.0687642591508294\\
9.1	0.0666525496389438\\
9.2	0.0645908683043217\\
9.3	0.0625792932133575\\
9.4	0.0606177715130224\\
9.5	0.0587061319888529\\
9.6	0.0568440967885719\\
9.7	0.0550312923476287\\
9.8	0.0532672595535146\\
9.9	0.0515514631858852\\
10	0.0498833006693581\\
10.1	0.0482621101754235\\
10.2	0.0466871781092397\\
10.3	0.0451577460162467\\
10.4	0.0436730169425371\\
10.5	0.0422321612818293\\
10.6	0.0408343221406967\\
10.7	0.0394786202524651\\
10.8	0.038164158468904\\
10.9	0.036890025857528\\
11	0.0356553014310112\\
11.1	0.0344590575339015\\
11.2	0.0333003629105252\\
11.3	0.0321782854766992\\
11.4	0.0310918948166135\\
11.5	0.0300402644250438\\
11.6	0.0290224737138704\\
11.7	0.0280376098007534\\
11.8	0.0270847690967198\\
11.9	0.0261630587083781\\
12	0.0252715976694741\\
12.1	0.0244095180155518\\
12.2	0.0235759657145772\\
12.3	0.022770101465522\\
12.4	0.0219911013760903\\
12.5	0.0212381575300001\\
12.6	0.0205104784535027\\
12.7	0.0198072894901378\\
12.8	0.0191278330920747\\
12.9	0.0184713690357816\\
13	0.0178371745691963\\
13.1	0.0172245444970333\\
13.2	0.0166327912103619\\
13.3	0.0160612446661206\\
13.4	0.0155092523217928\\
13.5	0.0149761790300602\\
13.6	0.0144614068978658\\
13.7	0.0139643351139627\\
13.8	0.0134843797486931\\
13.9	0.0130209735294311\\
14	0.0125735655948362\\
14.1	0.0121416212307976\\
14.2	0.0117246215907008\\
14.3	0.0113220634024181\\
14.4	0.0109334586642119\\
14.5	0.0105583343315408\\
14.6	0.0101962319965792\\
14.7	0.00984670756208763\\
14.8	0.00950933091111982\\
14.9	0.0091836855739044\\
15	0.00886936839310975\\
15.1	0.00856598918857607\\
15.2	0.00827317042248709\\
15.3	0.00799054686584987\\
15.4	0.00771776526705668\\
15.5	0.00745448402321434\\
15.6	0.00720037285484781\\
15.7	0.00695511248451014\\
15.8	0.00671839431976456\\
15.9	0.00648992014094263\\
16	0.00626940179402656\\
16.1	0.00605656088895262\\
16.2	0.00585112850358593\\
16.3	0.00565284489357489\\
16.4	0.00546145920825463\\
16.5	0.00527672921273455\\
16.6	0.00509842101627304\\
16.7	0.00492630880701452\\
16.8	0.00476017459313764\\
16.9	0.00459980795044121\\
17	0.00444500577637316\\
17.1	0.0042955720504896\\
17.2	0.00415131760131455\\
17.3	0.00401205987955618\\
17.4	0.00387762273762281\\
17.5	0.00374783621536986\\
17.6	0.00362253633199965\\
17.7	0.0035015648840267\\
17.8	0.00338476924921405\\
17.9	0.0032720021963794\\
18	0.00316312170096449\\
18.1	0.00305799076625658\\
18.2	0.00295647725014698\\
18.3	0.00285845369730869\\
18.4	0.00276379717667268\\
18.5	0.00267238912408062\\
18.6	0.00258411518999045\\
18.7	0.00249886509211069\\
18.8	0.00241653247283866\\
18.9	0.00233701476137828\\
19	0.00226021304041294\\
19.1	0.00218603191721021\\
19.2	0.00211437939903569\\
19.3	0.00204516677275462\\
19.4	0.0019783084885016\\
19.5	0.00191372204729971\\
19.6	0.0018513278925128\\
19.7	0.00179104930501616\\
19.8	0.00173281230197287\\
19.9	0.00167654553910546\\
20	0.00162218021635447\\
20.1	0.00156964998681778\\
20.2	0.00151889086886718\\
20.3	0.00146984116134052\\
20.4	0.00142244136171075\\
20.5	0.00137663408713502\\
20.6	0.00133236399828989\\
20.7	0.00128957772590075\\
20.8	0.00124822379987624\\
20.9	0.00120825258096079\\
21	0.00116961619482068\\
21.1	0.0011322684684816\\
21.2	0.00109616486903779\\
21.3	0.00106126244455545\\
21.4	0.00102751976709503\\
21.5	0.000994896877779733\\
21.6	0.000963355233839219\\
21.7	0.000932857657560121\\
21.8	0.00090336828707689\\
21.9	0.000874852528938619\\
22	0.000847277012389535\\
22.1	0.000820609545302827\\
22.2	0.00079481907170943\\
22.3	0.000769875630865306\\
22.4	0.000745750317802581\\
22.5	0.000722415245311716\\
22.6	0.000699843507303631\\
22.7	0.000678009143502463\\
22.8	0.000656887105421204\\
22.9	0.0006364532235742\\
23	0.000616684175881955\\
23.1	0.00059755745722528\\
23.2	0.000579051350107262\\
23.3	0.000561144896383004\\
23.4	0.00054381787001843\\
23.5	0.000527050750840842\\
23.6	0.000510824699245201\\
23.7	0.000495121531821384\\
23.8	0.000479923697868849\\
23.9	0.000465214256766413\\
24	0.000450976856165912\\
24.1	0.00043719571097967\\
24.2	0.000423855583132758\\
24.3	0.000410941762052081\\
24.4	0.000398440045865319\\
24.5	0.000386336723283718\\
24.6	0.000374618556143684\\
24.7	0.000363272762583009\\
24.8	0.000352287000828474\\
24.9	0.00034164935357236\\
25	0.00033134831291628\\
25.1	0.000321372765861481\\
25.2	0.000311711980325555\\
25.3	0.000302355591666198\\
25.4	0.000293293589693424\\
25.5	0.000284516306152246\\
25.6	0.000276014402658551\\
25.7	0.000267778859071519\\
25.8	0.000259800962286523\\
25.9	0.000252072295433075\\
26	0.000244584727462914\\
26.1	0.00023733040311393\\
26.2	0.000230301733236074\\
26.3	0.000223491385465994\\
26.4	0.000216892275237574\\
26.5	0.000210497557116038\\
26.6	0.000204300616443754\\
26.7	0.000198295061286289\\
26.8	0.00019247471466771\\
26.9	0.000186833607084515\\
27	0.000181365969287984\\
27.1	0.00017606622532512\\
27.2	0.000170928985828695\\
27.3	0.000165949041547299\\
27.4	0.000161121357106593\\
27.5	0.000156441064993318\\
27.6	0.000151903459753922\\
27.7	0.00014750399239995\\
27.8	0.000143238265012658\\
27.9	0.000139102025539581\\
28	0.000135091162776048\\
28.1	0.000131201701524901\\
28.2	0.000127429797927942\\
28.3	0.000123771734962835\\
28.4	0.000120223918099463\\
28.5	0.000116782871109933\\
28.6	0.000113445232026645\\
28.7	0.000110207749243071\\
28.8	0.000107067277752031\\
28.9	0.000104020775516521\\
29	0.000101065299968263\\
29.1	9.81980046293733e-05\\
29.2	9.54161358526829e-05\\
29.3	9.27170296764354e-05\\
29.4	9.00981087892253e-05\\
29.5	8.75568796012001e-05\\
29.6	8.50909294177005e-05\\
29.7	8.26979237116459e-05\\
29.8	8.03756034911149e-05\\
29.9	7.81217827586967e-05\\
30	7.59343460593177e-05\\
};
\end{axis}
\end{tikzpicture}%
    \caption{Probability density function of the particle diameter of NMC-622~\cite{Xuan2019} particles.}
    \label{fig:NMC_particle_diameter_distribution}
\end{figure}

The used material parameters are summarized in \cref{table:material_parameters}.
\begin{table}[ht]
    \renewcommand{\arraystretch}{1.2}
    \centering
    \caption{Material parameters.}
    \begin{tabular}{| c | c | c | c |}
  \hline
  \textbf{quantity}                                         & \textbf{NMC-622}                                                     & \textbf{lithium}                                                                                                           & \textbf{LPS}\\
  \hline
  bulk concentration ($c_\text{bulk}$)                      &  --                                                                  & --                                                                                                                         & $1.03 \cdot 10^4 \frac{\text{mol}}{\text{m}^3}$~\cite{Neumann2020} \\
  diffusion coefficient ($D$)                               & \cref{fig:NMC_transport_properties}~\cite{Neumann2020}               & --                                                                                                                         & -- \\
  electronic conductivity ($\sigma$)                        & \cref{fig:NMC_transport_properties}~\cite{Neumann2020}               & $10^5 \ \frac{\text{S}}{\text{m}}$~\cite{Neumann2020}                                                                      & -- \\
  exchange current density ($i_0$)                          & $4.98 \frac{\text{A}}{\text{m}^2}$~\cite{Schmidt2023}                & $8.87 \ \frac{\text{A}}{\text{m}^2}$ ~\cite{Neumann2020}                                                                   & -- \\
  growth law ($\mat{F}_\text{growth}$)                      & \cref{fig:NMC_OCP_growth}~\cite{Koerver2018}                         & \makecell{\cref{eq:growth_deposition} with \\ $p = \frac{M}{\rho} = 1.2998 \cdot 10^{-5} \ \frac{\text{m}^3}{\text{mol}}$} & --d \\
  ionic conductivity ($\sigma$)                             & --                                                                   & --                                                                                                                         & $1.2 \cdot 10^{-2} \ \frac{\text{S}}{\text{m}}$~\cite{Randau2020} \\
  lithiation range ($[\chi_\text{0\%}, \chi_\text{100\%}]$) & $[1, 0.404]$ defined                                                 & --                                                                                                                         & -- \\
  mass density ($\rho$)                                     & $5.03 \cdot 10^3 \ \frac{\text{kg}}{\text{m}^3}$~\cite{Schmidt2023}  & $0.53 \cdot 10^3 \frac{\text{kg}}{\text{m}^3}$~\cite{Lide2006}                                                             & $1.88 \cdot 10^3 \ \frac{\text{kg}}{\text{m}^3}$~\cite{Sakuda2013} \\
  max. concentration ($c_\text{max}$)                       & $5.19 \cdot 10^4 \ \frac{\text{mol}}{\text{m}^3}$~\cite{Neumann2020} & --                                                                                                                         & -- \\
  max. lithiation ($\chi_\text{max}$)                       & $1$~\cite{Neumann2020}                                               & --                                                                                                                         & -- \\
  open circuit voltage ($\Phi_0$)                           & \cref{fig:NMC_OCP_growth}~\cite{Kremer2019}                          & 0 V                                                                                                                        & -- \\
  \makecell{Particle diameter \\of powder ($D_\text{p}$)}                & \cref{fig:NMC_particle_diameter_distribution}~\cite{Song2021}        & --                                                                                                                         & -- \\
  Poisson's ratio ($\nu$)                                   & $0.3$~\cite{Xu2017}                                                  & $0.43$~\cite{Enghag2004}                                                                                                   & $0.27$~\cite{Yang2016} \\
  \makecell{transference number \\ of cations ($t_+$)}      & --                                                                   & --                                                                                                                         & $1$ (assumption)\\
  Young's modulus ($E$)                                     & $1.78 \cdot 10^{11} \ \text{Pa}$~\cite{Sun2017}                      & $4.9 \cdot 10^9 \ \text{Pa}$~\cite{Enghag2004}                                                                             & $2.89 \cdot 10^{10} \ \text{Pa}$~\cite{Yang2016} \\
  \hline
\end{tabular}
    \label{table:material_parameters}
\end{table}
The material parameters of the cathode active material that are modeled as a function of the lithiation state are shown in \cref{fig:NMC_lithiation_dependet_params}.
The size distribution of the NMC particles is shown in \cref{fig:NMC_particle_diameter_distribution}.
\section{Averaging elastic constants and the growth law}
\label{sec:averaging_mechanics}
The effective Young's modulus for a uniaxial deformation, i.e., one non-zero strain component of a homogeneous material, is obtained from Hooke's law
\begin{equation}
  \bar{E}_i = \frac{E_i}{1+\nu_i} + \frac{\nu_i E_i}{(1 - 2 \nu_i)(1 + \nu_i)}, 
\end{equation}
within an effective constitutive law $\sigma_{11} = \bar{E}_i \epsilon_{11}$. For averaging a composition of~$n$ components with share~$r_i$ and $\sum_i^n r_i = 1$, various strategies are conceivable. Arranging all components in parallel results in a homogenized Young's modulus of 
\begin{equation}
  \bar{E}_\text{par} = \sum_i^n r_i \bar{E}_i.
\end{equation}
Arranging all components in series results in
\begin{equation}
  \bar{E}_\text{ser} = \frac{1}{\sum_i^n \frac{r_i }{\bar{E}_i}}.
\end{equation}
A homogenized Young's modulus is now computed as the average of both extreme cases
\begin{equation}
  \bar{E} = \frac{\bar{E}_\text{par} + \bar{E}_\text{ser}}{2}.
\end{equation}
In a similar way, a homogenized growth law is derived. One component is considered to be volumetrically growing with the length after growth $l_\text{g} = l (1 + f)$ and the growth function~$f$, while the other component remains in its initial length $l_\text{o} = l$. In a parallel setup, a homogenized growth law is
\begin{equation}
  \bar{f}_\text{para} = \frac{\frac{f}{1+f} \bar{E}_\text{g} r_\text{g}}{\bar{E}_\text{o} r_\text{o} + \frac{\bar{E}_\text{g} r_\text{g}}{1 + f}}
\end{equation}
In an arrangement in series, the homogenized growth law is
\begin{equation}
  \bar{f}_\text{ser} = r_\text{g} f.
\end{equation}
The homogenized growth law is again obtained from averaging both cases, i.e.,
\begin{equation}
  \bar{f} = \frac{\bar{f}_\text{ser} + \bar{f}_\text{par}}{2}.
\end{equation}
\section{Quantification of the global sensitivity by Sobol indices}
\label{sec:sobol_indices}
Sobol indices quantify the influence of one parameter or a combination of parameters on an output quantity of a model or an experiment. The set of~$n$ input parameters is denoted by~$\vec{X}$, and the output quantity by~$Y(\vec{X})$. The expectation value~$E(Y)$ and the variance~$V(Y)$ of the output quantity are computed for different combinations of the input parameters. The Sobol indices of first-order are defined by the variance of the expectation value when keeping the parameter~$X_i$ fixed, i.e., $V_i = V(E(Y|X_i))$. Normalizing this value by the variance of the entire input parameter range results in the first-order Sobol index
\begin{equation}
    S_i = \frac{V(E(Y|X_i))}{V(Y)}.
\end{equation}
In a similar way, higher-order Sobol indices are defined. Then, the number of parameters that equal the order of the index is simultaneously fixed. The values of the higher order Sobol indices are then corrected by the values of the lower orders for these parameters, e.g., for the second-order Sobol index
\begin{equation}
    S_{ij} = \frac{V(E(Y|X_i,X_j))}{V(Y)} - S_i - S_j.
\end{equation}
By this definition, all Sobol indices up to $n$-th order sum up to one. \\
The total-order Sobol index is defined as the normalized expectation value of the variance if all parameters except one are fixed. It is defined by
\begin{equation}
    S^\text{t}_i = \frac{E(V(Y|X \backslash X_i))}{V(Y)}.
\end{equation}
By definition, all Sobol indices up to $n$-th order related to the parameter~$X_i$ sum up to~$S^\text{t}_i$. In particular, it tells if a parameter has an influence at all, either isolated or in combination with other parameters. \\
The numerical computation of the Sobol indices might become costly, as it requires computing the conditional expectation value and afterwards its variance. Typically, the computation of the expectation value and the variance is performed using Monte-Carlo integration~\cite{Sobol2001}. The computational costs for this nested operation to compute the first- and total-order Sobol indices scale with~$\mathcal{O}(M^2)$, where~$M$ is the number of Monte-Carlo samples that is recommended to be sufficiently large (c.f.~\cite{Wirthl2023}). The number of model evaluations can be reduced to~$M(2D+2)$ evaluations, with the number of input parameters~$D$ using an approach as done in~\cite{Saltelli2010}. The methodology is implemented in the software project \textit{QUEENS}~\cite{QUEENS} that has been used to perform the sensitivity analysis presented in this paper. 
\end{appendices}
\FloatBarrier
%
\bibliographystyle{IEEEtran}
\bibliography{literature}

\begin{thebibliography}{10}
\providecommand{\url}[1]{#1}
\csname url@samestyle\endcsname
\providecommand{\newblock}{\relax}
\providecommand{\bibinfo}[2]{#2}
\providecommand{\BIBentrySTDinterwordspacing}{\spaceskip=0pt\relax}
\providecommand{\BIBentryALTinterwordstretchfactor}{4}
\providecommand{\BIBentryALTinterwordspacing}{\spaceskip=\fontdimen2\font plus
\BIBentryALTinterwordstretchfactor\fontdimen3\font minus
  \fontdimen4\font\relax}
\providecommand{\BIBforeignlanguage}[2]{{%
\expandafter\ifx\csname l@#1\endcsname\relax
\typeout{** WARNING: IEEEtran.bst: No hyphenation pattern has been}%
\typeout{** loaded for the language `#1'. Using the pattern for}%
\typeout{** the default language instead.}%
\else
\language=\csname l@#1\endcsname
\fi
#2}}
\providecommand{\BIBdecl}{\relax}
\BIBdecl

\bibitem{Box1976}
G.~E.~P. Box, ``{S}cience and {S}tatistics,'' \emph{Journal of the American
  Statistical Association}, vol.~71, no. 356, pp. 791--799, Dec. 1976.

\bibitem{Bielefeld2023}
A.~Bielefeld, ``{H}ow to {D}evelop {U}seful {M}odels for {S}olid‐{S}tate
  {B}atteries – {A} {P}lea for {S}implicity and {I}nterdisciplinary
  {C}ooperation,'' \emph{Batteries \& Supercaps}, vol.~6, no.~9, Jul. 2023.

\bibitem{Ramadesigan2012}
V.~Ramadesigan, P.~W.~C. Northrop, S.~De, S.~Santhanagopalan, R.~D. Braatz, and
  V.~R. Subramanian, ``{M}odeling and {S}imulation of {L}ithium-{I}on
  {B}atteries from a {S}ystems {E}ngineering {P}erspective,'' \emph{Journal of
  The Electrochemical Society}, vol. 159, no.~3, pp. R31--R45, 2012.

\bibitem{Zhao2019a}
Y.~Zhao, P.~Stein, Y.~Bai, M.~Al-Siraj, Y.~Yang, and B.-X. Xu, ``A review on
  modeling of electro-chemo-mechanics in lithium-ion batteries,'' \emph{Journal
  of Power Sources}, vol. 413, pp. 259--283, 2019.

\bibitem{Krewer2018}
U.~Krewer, F.~Röder, E.~Harinath, R.~D. Braatz, B.~Bedürftig, and
  R.~Findeisen, ``{R}eview—{D}ynamic {M}odels of {L}i-{I}on {B}atteries for
  {D}iagnosis and {O}peration: {A} {R}eview and {P}erspective,'' \emph{Journal
  of The Electrochemical Society}, vol. 165, no.~16, pp. A3656--A3673, 2018.

\bibitem{Schmidt2023}
C.~P. Schmidt, S.~Sinzig, V.~Gravemeier, and W.~A. Wall, ``A three-dimensional
  finite element formulation coupling electrochemistry and solid mechanics on
  resolved microstructures of all-solid-state lithium-ion batteries,''
  \emph{Computer Methods in Applied Mechanics and Engineering}, vol. 417, p.
  116468, Dec. 2023.

\bibitem{Bucci2017}
G.~Bucci, T.~Swamy, S.~Bishop, B.~W. Sheldon, Y.-M. Chiang, and W.~C. Carter,
  ``{T}he {E}ffect of {S}tress on {B}attery-{E}lectrode {C}apacity,''
  \emph{Journal of The Electrochemical Society}, vol. 164, no.~4, pp.
  A645--A654, 2017.

\bibitem{Bistri2020}
D.~Bistri, A.~Afshar, and C.~V. Di~Leo, ``Modeling the chemo-mechanical
  behavior of all-solid-state batteries: a review.'' \emph{Meccanica}, vol.~56,
  no.~6, pp. 1523--1554, Jul. 2020.

\bibitem{Latz2015}
A.~Latz and J.~Zausch, ``Multiscale modeling of lithium ion batteries: thermal
  aspects,'' \emph{Beilstein Journal of Nanotechnology}, vol.~6, pp. 987--1007,
  2015.

\bibitem{Salvadori2015}
A.~Salvadori, D.~Grazioli, M.~G.~D. Geers, D.~Danilov, and P.~H.~L. Notten,
  ``{A} multiscale-compatible approach in modeling ionic transport in the
  electrolyte of ({L}ithium ion) batteries,'' \emph{Journal of Power Sources},
  vol. 293, pp. 892--911, 2015.

\bibitem{Bai2021}
Y.~Bai, D.~A. Santos, S.~Rezaei, P.~Stein, S.~Banerjee, and B.-X. Xu, ``A
  chemo-mechanical damage model at large deformation: numerical and
  experimental studies on polycrystalline energy materials,''
  \emph{International Journal of Solids and Structures}, vol. 228, p. 111099,
  2021.

\bibitem{Sinzig2024}
S.~Sinzig, C.~P. Schmidt, and W.~A. Wall, ``{A} {C}onservative and {E}fficient
  {M}odel for {G}rain {B}oundaries of {S}olid {E}lectrolytes in a {C}ontinuum
  {M}odel for {S}olid-{S}tate {B}atteries,'' \emph{Journal of The
  Electrochemical Society}, vol. 171, no.~4, p. 040505, 2024.

\bibitem{Sinzig2023}
S.~Sinzig, T.~Hollweck, C.~P. Schmidt, and W.~A. Wall, ``{A} {F}inite {E}lement
  {F}ormulation to {T}hree-{D}imensionally {R}esolve {S}pace-{C}harge {L}ayers
  in {S}olid {E}lectrolytes,'' \emph{Journal of The Electrochemical Society},
  vol. 170, no.~4, p. 040513, 2023.

\bibitem{Neumann2021}
A.~Neumann, T.~R. Hamann, T.~Danner, S.~Hein, K.~Becker-Steinberger,
  E.~Wachsman, and A.~Latz, ``{E}ffect of the {3D} {S}tructure and {G}rain
  {B}oundaries on {L}ithium {T}ransport in {G}arnet {S}olid {E}lectrolytes,''
  \emph{{ACS} Applied Energy Materials}, vol.~4, no.~5, pp. 4786--4804, 2021.

\bibitem{Braun2015}
S.~Braun, C.~Yada, and A.~Latz, ``Thermodynamically {C}onsistent {M}odel for
  {S}pace-{C}harge-{L}ayer {F}ormation in a {S}olid {E}lectrolyte,'' \emph{The
  Journal of Physical Chemistry C}, vol. 119, no.~39, pp. 22\,281--22\,288, sep
  2015.

\bibitem{Sinzig2023a}
S.~Sinzig, C.~P. Schmidt, and W.~A. Wall, ``{A}n {E}fficient {A}pproach to
  {I}nclude {T}ransport {E}ffects in {T}hin {C}oating {L}ayers in
  {E}lectrochemo-{M}echanical {M}odels for {A}ll-{S}olid-{S}tate {B}atteries,''
  \emph{Journal of The Electrochemical Society}, vol. 170, no.~10, p. 100532,
  2023.

\bibitem{Javed2020}
B.~Javed and M.~Koyama, ``{M}ulti-{P}hysics {S}imulation of {S}olid-{S}tate
  {B}atteries with {A}ctive {M}aterial {C}oating,'' \emph{Journal of The
  Electrochemical Society}, vol. 167, no.~2, p. 020521, jan 2020.

\bibitem{Schmidt2024}
C.~P. Schmidt, S.~Sinzig, and W.~A. Wall, ``An electro-chemo-mechanic model
  resolving delamination between components in complex microstructures of
  solid-state batteries,'' 2024.

\bibitem{Fang2022}
R.~Fang, C.~P. Schmidt, and W.~A. Wall, ``A coupled finite element approach to
  spatially resolved lithium plating and stripping in three-dimensional anode
  microstructures of lithium-ion cells,'' \emph{Journal of Computational
  Physics}, p. 111179, 2022.

\bibitem{Hein2020a}
S.~Hein, T.~Danner, and A.~Latz, ``{A}n {E}lectrochemical {M}odel of {L}ithium
  {P}lating and {S}tripping in {L}ithium {I}on {B}atteries,'' \emph{{ACS}
  Applied Energy Materials}, vol.~3, no.~9, pp. 8519--8531, 2020.

\bibitem{Horstmann2019}
B.~Horstmann, F.~Single, and A.~Latz, ``Review on multi-scale models of
  solid-electrolyte interphase formation,'' \emph{Current Opinion in
  Electrochemistry}, vol.~13, pp. 61--69, 2019.

\bibitem{Jokar2016}
A.~Jokar, B.~Rajabloo, M.~D{\'{e}}silets, and M.~Lacroix, ``{R}eview of
  simplified {P}seudo-two-{D}imensional models of lithium-ion batteries,''
  \emph{Journal of Power Sources}, vol. 327, pp. 44--55, 2016.

\bibitem{Traskunov2022}
I.~Traskunov and A.~Latz, ``Novel local fluctuation-preserving upscaling
  techniques for heterogeneous reaction-transport equation in porous media,''
  \emph{Electrochimica Acta}, vol. 434, p. 141248, Dec. 2022.

\bibitem{Kirk2021}
T.~L. Kirk, C.~P. Please, and S.~Jon~Chapman, ``{P}hysical {M}odelling of the
  {S}low {V}oltage {R}elaxation {P}henomenon in {L}ithium-{I}on {B}atteries,''
  \emph{Journal of The Electrochemical Society}, vol. 168, no.~6, p. 060554,
  Jun. 2021.

\bibitem{Safari2009}
M.~Safari, M.~Morcrette, A.~Teyssot, and C.~Delacourt, ``{M}ultimodal
  {P}hysics-{B}ased {A}ging {M}odel for {L}ife {P}rediction of {L}i-{I}on
  {B}atteries,'' \emph{Journal of The Electrochemical Society}, vol. 156,
  no.~3, p. A145, 2009.

\bibitem{Yang2018b}
X.-G. Yang, S.~Ge, T.~Liu, Y.~Leng, and C.-Y. Wang, ``A look into the voltage
  plateau signal for detection and quantification of lithium plating in
  lithium-ion cells,'' \emph{Journal of Power Sources}, vol. 395, pp. 251--261,
  Aug. 2018.

\bibitem{Khalik2021}
Z.~Khalik, M.~Donkers, J.~Sturm, and H.~Bergveld, ``{P}arameter estimation of
  the {D}oyle–{F}uller–{N}ewman model for {L}ithium-ion batteries by
  parameter normalization, grouping, and sensitivity analysis,'' \emph{Journal
  of Power Sources}, vol. 499, p. 229901, Jul. 2021.

\bibitem{Lu2022}
D.~Lu, M.~Scott~Trimboli, G.~Fan, Y.~Wang, and G.~L. Plett, ``{N}ondestructive
  {EIS} {T}esting to {E}stimate a {S}ubset of {P}hysics-based-model {P}arameter
  {V}alues for {L}ithium-ion {C}ells,'' \emph{Journal of The Electrochemical
  Society}, vol. 169, no.~8, p. 080504, Aug. 2022.

\bibitem{An2021}
F.~An, W.~Zhou, and P.~Li, ``A comparison of model prediction from {P2D} and
  particle packing with experiment,'' \emph{Electrochimica Acta}, vol. 370, p.
  137775, Feb. 2021.

\bibitem{Schmidt2021}
A.~Schmidt, E.~Ramani, T.~Carraro, J.~Joos, A.~Weber, M.~Kamlah, and
  E.~Ivers-Tiffée, ``{U}nderstanding {D}eviations between {S}patially
  {R}esolved and {H}omogenized {C}athode {M}odels of {L}ithium‐{I}on
  {B}atteries,'' \emph{Energy Technology}, vol.~9, no.~6, Jan. 2021.

\bibitem{Goldin2012}
G.~M. Goldin, A.~M. Colclasure, A.~H. Wiedemann, and R.~J. Kee,
  ``{T}hree-dimensional particle-resolved models of {L}i-ion batteries to
  assist the evaluation of empirical parameters in one-dimensional models,''
  \emph{Electrochimica Acta}, vol.~64, pp. 118--129, 2012.

\bibitem{Park2015}
J.~Park, S.~Lee, J.~Hoffmann, and A.~M. Sastry, ``{S}tudy of the {E}ffect of
  {E}lectrode {M}icrostructures on {B}attery {P}erformance,'' in \emph{Volume
  6A: Energy}, ser. IMECE2015.\hskip 1em plus 0.5em minus 0.4em\relax American
  Society of Mechanical Engineers, Nov. 2015.

\bibitem{Doyle1992}
M.~Doyle, T.~F. Fuller, and J.~Newman, ``{M}odeling of {G}alvanostatic {C}harge
  and {D}ischarge of the {L}ithium/{P}olymer/{I}nsertion {C}ell,''
  \emph{Journal of The Electrochemical Society}, vol. 140, no.~6, pp.
  1526--1533, Jun. 1993.

\bibitem{Doyle1995}
M.~Doyle and J.~Newman, ``The use of mathematical modeling in the design of
  lithium/polymer battery systems,'' \emph{Electrochimica Acta}, vol.~40, no.
  13–14, pp. 2191--2196, Oct. 1995.

\bibitem{Pereira2022}
D.~J. Pereira, A.~M. Aleman, J.~W. Weidner, and T.~R. Garrick, ``{A}
  {M}echano-{E}lectrochemical {B}attery {M}odel that {A}ccounts for
  {P}referential {L}ithiation {I}nside {B}lended {S}ilicon {G}raphite
  ({S}i/{C}) {A}nodes,'' \emph{Journal of The Electrochemical Society}, vol.
  169, no.~2, p. 020577, 2022.

\bibitem{Neumann2020}
A.~Neumann, S.~Randau, K.~Becker-Steinberger, T.~Danner, S.~Hein, Z.~Ning,
  J.~Marrow, F.~H. Richter, J.~Janek, and A.~Latz, ``Analysis of {I}nterfacial
  {E}ffects in {A}ll-{S}olid-{S}tate {B}atteries with {T}hiophosphate {S}olid
  {E}lectrolytes,'' \emph{{ACS} Applied Materials {\&} Interfaces}, vol.~12,
  no.~8, pp. 9277--9291, 2020.

\bibitem{Wirthl2023}
B.~Wirthl, S.~Brandstaeter, J.~Nitzler, B.~A. Schrefler, and W.~A. Wall,
  ``Global sensitivity analysis based on {G}aussian-process metamodelling for
  complex biomechanical problems,'' \emph{International Journal for Numerical
  Methods in Biomedical Engineering}, vol.~39, no.~3, jan 2023.

\bibitem{4C}
{4C}, ``{4C}: A {C}omprehensive {M}ulti-{P}hysics {S}imulation {F}ramework,''
  \url{https://www.4c-multiphysics.org}, 2024, accessed: 7 May 2024.

\bibitem{Nitzler2022}
J.~Nitzler, J.~Biehler, N.~Fehn, P.-S. Koutsourelakis, and W.~A. Wall, ``A
  generalized probabilistic learning approach for multi-fidelity uncertainty
  quantification in complex physical simulations,'' \emph{Computer Methods in
  Applied Mechanics and Engineering}, vol. 400, p. 115600, Oct. 2022.

\bibitem{Koerver2018}
R.~Koerver, W.~Zhang, L.~de~Biasi, S.~Schweidler, A.~O. Kondrakov, S.~Kolling,
  T.~Brezesinski, P.~Hartmann, W.~G. Zeier, and J.~Janek, ``Chemo-mechanical
  expansion of lithium electrode materials {\textendash} on the route to
  mechanically optimized all-solid-state batteries,'' \emph{Energy {\&}
  Environmental Science}, vol.~11, no.~8, pp. 2142--2158, 2018.

\bibitem{Randau2020}
S.~Randau, D.~A. Weber, O.~Kötz, R.~Koerver, P.~Braun, A.~Weber,
  E.~Ivers-Tiff{\'{e}}e, T.~Adermann, J.~Kulisch, W.~G. Zeier, F.~H. Richter,
  and J.~Janek, ``Benchmarking the performance of all-solid-state lithium
  batteries,'' \emph{Nature Energy}, vol.~5, no.~3, pp. 259--270, mar 2020.

\bibitem{Lide2006}
D.~Lide, \emph{{CRC} handbook of chemistry and physics: a ready-reference book
  of chemical and physical data}.\hskip 1em plus 0.5em minus 0.4em\relax Boca
  Raton, Fla: CRC Taylor \& Francis, 2006.

\bibitem{Sakuda2013}
A.~Sakuda, A.~Hayashi, and M.~Tatsumisago, ``{S}ulfide {S}olid {E}lectrolyte
  with {F}avorable {M}echanical {P}roperty for {A}ll-{S}olid-{S}tate {L}ithium
  {B}attery,'' \emph{Scientific Reports}, vol.~3, no.~1, 2013.

\bibitem{Kremer2019}
L.~S. Kremer, A.~Hoffmann, T.~Danner, S.~Hein, B.~Prifling, D.~Westhoff,
  C.~Dreer, A.~Latz, V.~Schmidt, and M.~Wohlfahrt-Mehrens, ``{M}anufacturing
  {P}rocess for {I}mproved {U}ltra-{T}hick {C}athodes in {H}igh-{E}nergy
  {L}ithium-{I}on {B}atteries,'' \emph{Energy Technology}, vol.~8, no.~2, p.
  1900167, 2019.

\bibitem{Song2021}
Z.~Song, P.~Zhu, W.~Pfleging, and J.~Sun, ``Electrochemical {P}erformance of
  {T}hick-{F}ilm {L}i({N}i$_{0.6}${M}n$_{0.2}${C}o$_{0.2}$){O}$_2$ {C}athode
  with {H}ierarchic {S}tructures and {L}aser {A}blation,''
  \emph{Nanomaterials}, vol.~11, no.~11, p. 2962, Nov. 2021.

\bibitem{Xu2017}
R.~Xu, H.~Sun, L.~S. de~Vasconcelos, and K.~Zhao, ``{M}echanical and
  {S}tructural {D}egradation of {L}i{N}i$_x${M}n$_y${C}o$_z${O}$_2$ {C}athode
  in {L}i-{I}on {B}atteries: {A}n {E}xperimental {S}tudy,'' \emph{Journal of
  The Electrochemical Society}, vol. 164, no.~13, pp. A3333--A3341, 2017.

\bibitem{Enghag2004}
P.~Enghag, \emph{Encyclopedia of the elements}.\hskip 1em plus 0.5em minus
  0.4em\relax Weinheim, Germany: Wiley-VCH Verlag, Aug. 2004.

\bibitem{Yang2016}
Y.~Yang, Q.~Wu, Y.~Cui, Y.~Chen, S.~Shi, R.-Z. Wang, and H.~Yan, ``{E}lastic
  {P}roperties, {D}efect {T}hermodynamics, {E}lectrochemical {W}indow, {P}hase
  {S}tability, and {L}i$^+$ {M}obility of {L}i$_3${PS}$_4$: {I}nsights from
  {F}irst-{P}rinciples {C}alculations,'' \emph{{ACS} Applied Materials {\&}
  Interfaces}, vol.~8, no.~38, pp. 25\,229--25\,242, 2016.

\bibitem{Sun2017}
H.~Sun and K.~Zhao, ``{E}lectronic {S}tructure and {C}omparative {P}roperties
  of {L}i{N}i$_x${M}n$_y${C}o$_z${O}$_2$ {C}athode {M}aterials,'' \emph{The
  Journal of Physical Chemistry C}, vol. 121, no.~11, pp. 6002--6010, 2017.

\bibitem{Xuan2019}
W.~Xuan, A.~Otsuki, and A.~Chagnes, ``Investigation of the leaching mechanism
  of {NMC} 811 ({L}i{N}i$_{0.8}${M}n$_{0.1}${C}o$_{0.1}${O}$_2$) by
  hydrochloric acid for recycling lithium ion battery cathodes,'' \emph{RSC
  Advances}, vol.~9, no.~66, pp. 38\,612--38\,618, 2019.

\bibitem{Sobol2001}
I.~Sobol, ``Global sensitivity indices for nonlinear mathematical models and
  their {M}onte {C}arlo estimates,'' \emph{Mathematics and Computers in
  Simulation}, vol.~55, no. 1–3, pp. 271--280, Feb. 2001.

\bibitem{Saltelli2010}
A.~Saltelli, P.~Annoni, I.~Azzini, F.~Campolongo, M.~Ratto, and S.~Tarantola,
  ``Variance based sensitivity analysis of model output. {D}esign and estimator
  for the total sensitivity index,'' \emph{Computer Physics Communications},
  vol. 181, no.~2, pp. 259--270, Feb. 2010.

\bibitem{QUEENS}
QUEENS, ``{A} general purpose framework for {U}ncertainty {Q}uantification,
  {P}hysics-{I}nformed {M}achine {L}earning, {B}ayesian {O}ptimization,
  {I}nverse {P}roblems and {S}imulation {A}nalytics on distributed computer
  systems, accessed: {M}arch 19, 2024.'' URL
  \url{https://queens_community.pages.gitlab.lrz.de/website/}.

\end{thebibliography}
\end{document}